\documentclass[letterpaper,10pt]{article}
\pdfoutput=1
\usepackage{graphicx,times}
\usepackage{amssymb,amsmath}
\usepackage{pdfpages}
\usepackage[colorlinks=true, urlcolor=dablue, linkcolor=blue, citecolor=blue, pdftex]{hyperref}
\hypersetup{pdfstartview={FitV}, pdfpagemode = UseNone, pdfpagelayout=SinglePage, bookmarksnumbered={true}}
\newcommand{\mean}[1]{\langle #1 \rangle}%
\newcommand{\mypart}[1]{}
\newcommand{\comb}[1]{[ #1 ]}%

\def\og{\hat{\omega}_g}
\def\obare{\hat{\omega}}

\def\hg{h_g}
\def\sg{\sum_{g}}
\def\emuhg{e^{\mu \hg}}
\def\Xg{\hat{X}_g}
\def\Ham{\hat{{\cal H}}}
\def\Heff{{\hat{{\cal H}}}_{\text{eff}}}
\def\ng{\mathrm{n}(g)}
\def\dg{\mathrm{dg}(g)}
\newcommand{\dmu}[1][]{\partial_\mu {#1}}
\newcommand{\Ov}[1][]{\hat{{\cal O}}^{#1}}
\newcommand{\Z}[1][]{\hat{Z}^{#1}}
\newcommand{\cZ}[1][]{\hat{z}^{(#1)}}
\newcommand{\Zc}[1][]{\hat{z}_{c}^{(#1)}}
\newcommand{\1}[1][]{\hat{1}^{#1}}
\newcommand{\incl}[1]{\mypart{#1}}
\newcommand{\Set}[2][]{{\cal S}_{#1}^{#2}}

\newcommand{\muzero}[1][]{{#1} \rvert_{\mu = 0}}
\newcommand{\muzerobetaone}[1][]{{#1} \rvert_{\substack{\mu = 0 \\ \beta=1}}}

\def\logZ{\ln \Z}
\def\dlogZ{\dmu[\logZ]}

\def\eqmuzerobetaone{\underset{\substack{\mu = 0 \\ \beta=1}}{=}}
\newcommand{\commute}[2][]{\left [ {#1} , {#2} \right ]}
\newcommand{\halfcommute}[1][]{\left [ {#1} , \right .}
\newcommand{\acommute}[2][]{\left \{ {#1} , {#2} \right \} }
\newcommand{\halfacommute}[1][]{\left \{ {#1} , \right . }
\newcommand{\ket}[1]{|{#1}\rangle}
\newcommand{\bra}[1]{\langle{#1}|}
\newcommand{\sca}[2]{\langle{#1}|{#2}\rangle}

\begin{document}
\incl{%
\author{David~Schwandt}
\author{Matthieu~Mambrini}
\author{Didier~Poilblanc}
\affiliation{Laboratoire de Physique Th\'eorique, CNRS \& Universit\'e de Toulouse,
 F-31062 Toulouse, France}
\date{\today}
\title{Generalized Hardcore Dimer Models approach to low-energy Heisenberg frustrated antiferromagnets: general properties and application to the kagome antiferromagnet}
\begin{abstract}
We propose a general non-perturbative scheme that quantitatively maps the low-energy sector of spin-$1/2$ frustrated Heisenberg antiferromagnets to effective Generalized Quantum Dimer Models. We develop the formal lattice independent frame and establish some important results on (i) the locality of the generated Hamiltonians (ii) how full resummations can be performed in this renormalization scheme. The method is then applied to the much debated {\it kagome} antiferromagnet for which a fully resummed effective Hamiltonian -- shown to capture the essential properties and provide deep insights on the microscopic model [D. Poilblanc, M. Mambrini and D. Schwandt, arXiv:0912.0724] -- is derived.
\end{abstract}
\section{Introduction}
\label{sec:introduction}

After decades of theoretical efforts, understanding the low-energy properties of bidimensionnal frustrated Quantum Heisenberg Antiferromagnets (QHAF) is still a notoriously puzzling problem. In contrast to  unfrustated models where powerful paradigms have emerged to describe both the magnetically ordered ground state (N\'eel state) and the excitations (magnons), a general theoretical framework is still lacking. Not only the low-energy physics is in general poorly characterized -- one can have in mind the lack of consensus on archetypal models such as $J_1-J_2$ or {\it kagome} antiferromagnets -- but generic tools to identify the low energy degrees of freedom and understand their effective interaction are missing~\cite{j1j2}.
Recent discoveries of new materials such as {\it herbertsmithite}~\cite{herbertsmithite}, believed to be almost perfect spin-1/2 kagome antiferromagnets, have revived the former interest for bidimensionnal frustrated QHAF 
and call more than ever for novel theoretical approaches that could provide a quantitative understanding of these materials and their exotic low-temperature
non-magnetic phases~\cite{j1j2}. 

The underlying hardness of the problem comes from the fact that the usual theoretical frame for treating such systems typically requires to (i) identify a well controlled limit (classical limit, unperturbed limit, \ldots) and then (ii) introduce quantum fluctuations and/or corrections to this limit in a putatively perturbative way. This scheme however fails for bidimensionnal frustrated QHAF since none of the above mentioned points is fulfilled : (i) because of frustration the classical limit is highly degenerate and often not fully understood, (ii) quantum fluctuations generally do not act as a small perturbation on the classical limit. As a direct consequence, the low-energy physics of these systems is characterized by very exotic magnetically disordered states such as 
Resonating Valence Bond (RVB) states~\cite{anderson}, Valence Bond Crystals (VBC)~\cite{vbc} or spin liquids~\cite{lee-et-al}.

The aim of this article is to build a {\em non-perturbative} scheme allowing to derive an effective low-energy Generalized Quantum Dimer Model (GQDM) that captures the physics of the microscopic model. In turn, the GQDM can be used to efficiently investigate the low-energy properties of the microscopic Hamiltonian either by numerical techniques on large systems (by Exact Diagonalizations or Quantum Monte Carlo when the GQDM has no sign problem) or by analytical techniques such as gauge theory mapping.

Basically, the method relies on (i) the identification of a versatile and relevant manifold of states at low-energy and (ii) the projection of the microscopic Hamiltonian in this manifold. Such a framework, first initiated by Rokhsar and Kivelson~\cite{rokhsar} in a different context, has been put here 
on a formal basis and greatly extended, providing a novel versatile and systematic expansion scheme to deal with frustrated magnets.

The paper is organized as follows: in the first part we present the derivation scheme focusing on important results and highlighting practical implementation rules. To increase readability, we make an extensive use of appendices in which all technical details and demonstrations are postponed. The second part is devoted to the application of the method to the kagome antiferromagnet. We derive the parameter-free GQDM Hamiltonian that has been proved to give deeper insights to the low energy physics of the kagome antiferromagnet in a companion paper\cite{qdmkagome}. Then we discuss its key properties focusing on the fact that this untuned GQDM model lies at the vicinity of several competing exotic phases, a $\mathbb{Z}_2$ dimer liquid and two VBC phases, sheding light on the critical properties of the microscopic model. 
For completeness, we provide complementary results for the Kagome lattice (see Supplementary material at the end of the article).

\section{Effective Hamiltonian mapping scheme}
\label{sec:method}
\subsection{Low energy manifold}

In the past decades numerical studies played a central role in establishing unbiased and reliable results on bidimensionnal frustrated QHAF\cite{j1j2}. More precisely, Exact Diagonalizations (ED) of the Heisenberg Hamiltonian on finite and generally small clusters were able to provide precise insights on these problems. Other numerical methods, such as Quantum Monte Carlo (QMC) or Density Matrix Renormalization Group (DMRG), have proven their relative inadequacy to treat such systems and failed to be generalized for reasons that appear more and more to be fundamental rather than purely technical.

Based on ED results, two salient facts can be retained to draw a phenomenological low-energy picture: bidimensionnal frustrated QHAF (i) remain magnetically disordered at zero temperature (i.e. the zero temperature spin-spin correlation function $\langle \hat{\mathbf{S}}_i.\hat{\mathbf{S}}_j \rangle $ does not display long range antiferromagnetic order), (ii) do not break the SU(2) symmetry, the groundstate and low-lying excitations are singlets (i.e. eigenstates of the total spin $\hat{\mathbf{S}}$ with a zero eigenvalue).

Singlet states are intimately related to Valence Bond (VB) states, namely products over pairs of spins of arbitrary range SU(2) singlet wavefunctions (also called {\it SU(2) dimers} in the following). Indeed, any singlet state can be expressed as a linear superposition of VB states\cite{rumer,hulthen}. It is therefore tempting to build a low-energy singlet theory by representing the original spin Hamiltonian in the VB basis rather than using spin variables. However, the set of all VB states is massively overcomplete, as the number of VB states is much larger than the number of singlets. This originates from the fact that in the VB representation dimers are allowed to connect arbitrarily distant sites. On the other hand, Liang, Dou\c{c}ot and Anderson showed that a deep connection exists between the spin correlation length of a singlet wavefunction and its bond-length distribution\cite{liang}: a finite correlation length for $\langle \hat{\mathbf{S}}_i.\hat{\mathbf{S}}_j \rangle $ can only be obtained for a sufficiently fast decay with dimer length of dimer amplitudes in the wavefunction.

This strongly suggests that a versatile framework for describing the magnetically disordered low energy manifold of bidimensionnal frustrated QHAF can be obtained by considering only the restriction of the VB states to short range or Nearest Neighbor Valence Bond (NNVB) states. The precise numerical check of this statement as well as the ability of the NNVB set of states to capture the low energy physics of bidimensionnal frustrated QHAF has been performed in previous work in the case of the $J_1-J_2-J_3$ model on a square lattice\cite{J1J2J3NNVB} and for the kagome antiferromagnet\cite{kagomeNNVB,kagomeNNVBimpurity}.

As a concluding remark let us mention that, contrary to the unrestricted VB set of states, the NNVB states are linearly independent for most low-connectivity lattices. This can be numerically checked for reasonably large systems on the square, triangular and kagome lattices\cite{mambriniunp} and has been recently proved for the kagome lattice\cite{seidel}. However, as the number of NNRVB states on those lattices is smaller than the total number of singlets, it is clear that this restriction further reduces the singlet Hilbert space which is a key advantage for numerical computations.  

\subsection{Effective Hamiltonian}

For convenience, most of the illustrations provided in this section are represented using the kagome lattice. However, the formalism is general and all technical details, provided in the Appendices \ref{subsec:signs} to \ref{sec:fullyconnected} and to which we extensively refer, use a lattice-independent formulation.

\subsubsection{Linear algebra problem}

{\it Non-orthogonality}. A crucial property of VB states is their non-orthogonality: for any $\vert \varphi \rangle$ and $\vert \psi \rangle$, the scalar product $\sca{\varphi}{\psi}$ is always non-zero and, as recalled in Appendix \ref{subsec:signs},
\begin{equation}
\label{eq:nonorth}
\sca{\varphi}{\psi} = \alpha^{N-2 n_l \left ( \varphi,\psi \right )},
\end{equation}
where $N$ is the number of sites of the system and $n_l \left ( \varphi,\psi \right )$ is the number of loops in the overlap graph (see figure \ref{fig:overlap}). In the general case, the overlap $\sca{\varphi}{\psi}$ is a signed quantity. This is a direct consequence of the SU(2) dimer wavefunction antisymmetry which requires in turn to specify a conventional orientation of lattice bonds. 

When restricting to NNVB states, this choice  is constrained by the nature of the lattice. The {\it bosonic} convention requires a prescription of bond orientations and hence is more adapted to bipartite lattices. On the other hand, the {\it fermionic} convention, being free of any dimer ordering, is convenient for non-bipartite lattices (see Appendix \ref{subsec:signs}). Using the fermionic convention (for arbitrary lattices) or the bosonic convention on bipartite lattices allow to absorb the sign of $\sca{\varphi}{\psi}$ in the single parameter $\alpha$ by taking $\alpha_b = 1 / \sqrt{2}$ (respectively $\alpha_f = i / \sqrt{2}$).

\begin{figure}[!h]
\begin{center}
\includegraphics[width=\linewidth]{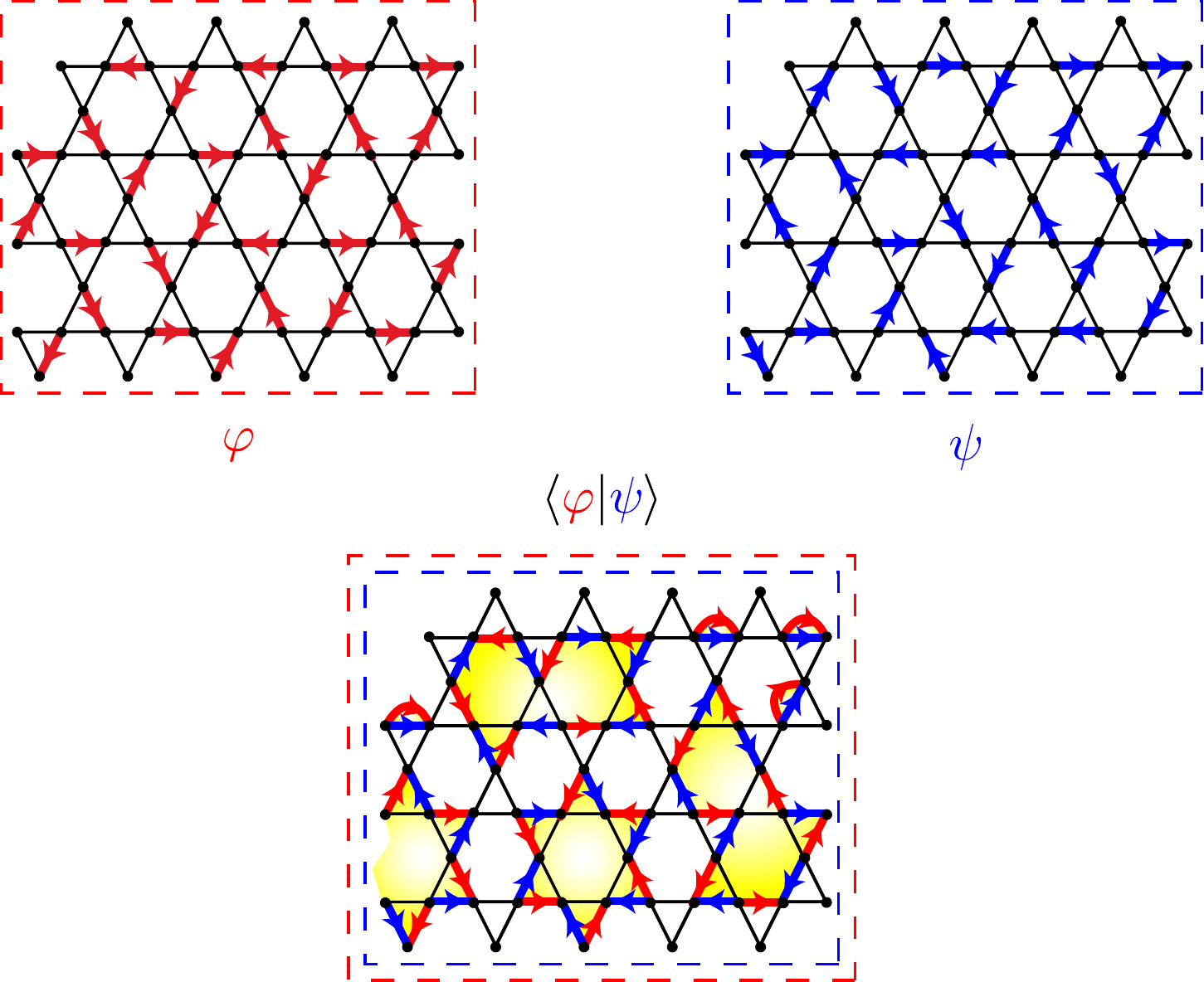}
\end{center}
\caption{\label{fig:overlap} (Color online) Overlap $\sca{\varphi}{\psi}$ between two VB wave functions $\ket{\varphi}$ and $\ket{\psi}$. The amplitude of $\sca{\varphi}{\psi}$ is driven by the number of loops appearing in the overlap graph (bottom frame).}
\end{figure}

{\it Dual basis}. When working with a non-orthogonal basis $\{\ket{\psi}\}$ as the NNVB states, it is generally useful to introduce its dual basis $\{\ket{\psi^{\star}}\}$, such that
\begin{equation}
\sca{\varphi^{\star}}{\psi}=\delta_{\varphi,\psi}.
\end{equation}
Of course, both span the same space, but the dual states are clearly not VB states, as the overlap between VB states is always non-zero. As we want to work with the NNVB basis, one can then further introduce the operator
\begin{equation}
\label{eq:oclosing}
\hat{\cal O}=\sum_\chi\ket{\chi}\bra{\chi},
\end{equation}
that transforms the dual states into NNVB states: $\hat{\cal O}\ket{\psi^\star}=\ket{\psi}$. Its inverse is given by 
\begin{equation}
\label{eq:invoclosing}
\hat{\cal O}^{-1}=\sum_\chi\ket{\chi^\star}\bra{\chi^\star}
\end{equation}
and transforms the NNVB states back to the dual states. Therefore $\langle \varphi \vert \hat{\cal O}^{-1} \vert \psi \rangle=\sca{\varphi^{\star}}{\psi}=\delta_{\varphi,\psi}$ and $\{\ket{\psi^\perp}\}=\{\hat{\cal O}^{-1/2}\ket{\psi}\}$ is easily seen to be an orthonormal basis\cite{rokhsar}.

In order to calculate matrix elements in a non-orthogonal basis, one may define
\begin{equation}
\begin{split}
{\cal A}_{\varphi,\psi} &= \langle \varphi^\star \vert \hat{\cal A} \vert \psi \rangle \\
												&=\langle \varphi \vert \hat{\cal O}^{-1}\hat{\cal A} \vert \psi \rangle.
\end{split}
\end{equation}
Setting $\hat{\cal H}=\hat{\cal O}\hat{\cal H}_H$, we find
\begin{align}
{\cal O}_{\varphi,\psi} & =\sca{\varphi}{\psi},\\
{\cal H}_{\varphi,\psi} & =\langle \varphi \vert \hat{\cal H}_H \vert \psi \rangle.
\end{align}
${\cal O}_{\varphi,\psi}$ is called \textit{overlap} matrix and the \textit{Heisenberg} Hamiltonian for bidimensionnal frustrated QHAF in its general form is given by
\begin{equation}
\hat{\cal H}_H = \sum_{i,j} J_{i,j} \hat{\mathbf{S}}_i.\hat{\mathbf{S}}_j.
\end{equation}

{\it Matrix elements}. The matrix elements ${\cal H}_{\varphi,\psi}$ can be obtained from the overlap matrix elements ${\cal O}_{\varphi,\psi}$ straightforwardly, generalizing Ref.~\onlinecite{sutherland}. It can be shown (see Appendix \ref{subsec:signs} and figure \ref{fig:boucles} for details) that 
\begin{equation}
\label{eq:Hamov}
\langle \varphi \vert \hat{\cal H}_H \vert \psi \rangle=\sum_{i,j} J_{i,j}\ \varepsilon_{i,j}^{\varphi,\psi} \sca{\varphi}{\psi},
\end{equation}
where $\varepsilon_{i,j}^{\varphi,\psi}$ can be derived from the overlap graph. Figure (\ref{fig:boucles}) illustrates the fact that $\varepsilon_{i,j}^{\varphi,\psi}\in\{0,+3/4,-3/4\}$. In the case where $i$ and $j$ are on two distinct loops
in the overlap graph of $\ket{\varphi}$ and $\ket{\psi}$, we have $\varepsilon_{i,j}^{\varphi,\psi}=0$. When $i$ and $j$ are on the same loop one finds that $\varepsilon_{i,j}^{\varphi,\psi}=+3/4$ ($\varepsilon_{i,j}^{\varphi,\psi}=-3/4$) if the distance between $i$ and $j$ is even (odd) with respect to the loop formed by $\ket{\varphi}$ and $\ket{\psi}$. 

\begin{figure}
\begin{center}
\includegraphics[width=0.95\linewidth]{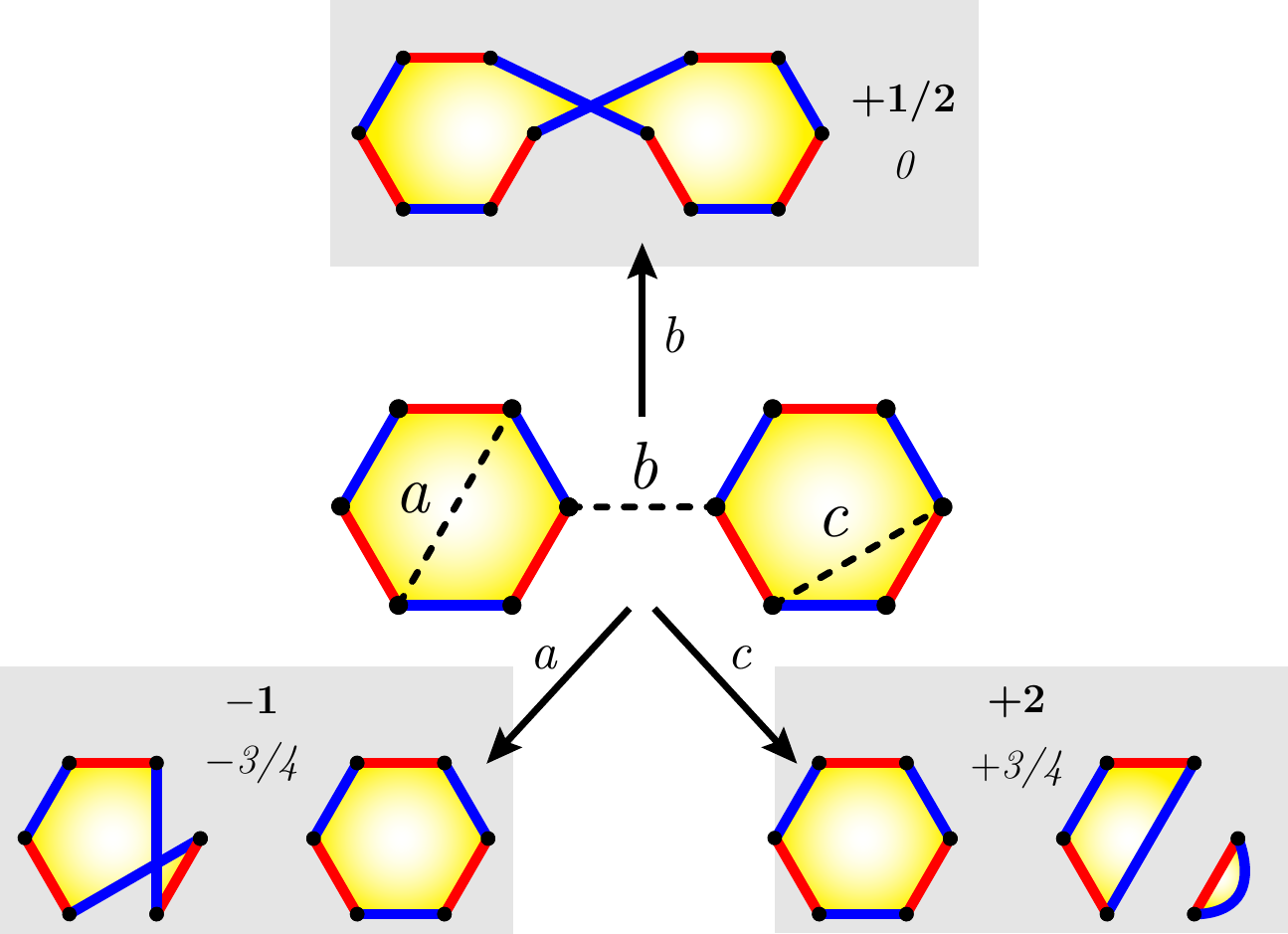}
\end{center}
\caption{\label{fig:boucles} (Color online) Application of $\hat{P}_l = 2 \hat{\mathbf{S}}_i.\hat{\mathbf{S}}_j +1/2$ on a loop diagram $\langle \varphi \vert \chi \rangle$ for the three types of bonds $l=(i,j)$ described in the text: ``internal odd'' ({\it a}), ``external'' ({\it b}) and ``internal even'' ({\it c}). The state $\vert \chi \rangle$ (respectively $\vert \varphi  \rangle$) is represented with blue (respectively red) bonds. Note that $P_l$ is applied on $\vert \chi \rangle$. Figures nearby the diagrams are the ratios $\langle \varphi \vert \hat{P}_l \vert \chi \rangle / \langle \varphi \vert \chi \rangle$ (bold) and $\langle \varphi \vert (4/3)\hat{\mathbf{S}}_i.\hat{\mathbf{S}}_j \vert \chi \rangle / \langle \varphi \vert \chi \rangle$ (italic).}
\end{figure}

In the case where $J_{i,j}$ is uniform for nearest neighbors pairs $\langle i,j \rangle$ of sites $i$ and $j$, we introduce the nearest neighbor coupling $J_1=J_{\langle i,j \rangle }$. Denoting the net length of all non-trivial loops by ${\cal L}_{\varphi,\psi}$, the number of trivial (i.e. length-2) loops is given by $N/2-{\cal L}_{\varphi,\psi}/2$. For convenience we will shift away all contributions from trivial loops and rescale the Hamiltonian by a factor $4/3$, and thus from now on we replace $\hat{\cal H}_H$ by $(4/3) \hat{\cal H}_H +J_1 N/2$.

In this new convention we have $\varepsilon_{i,j}\in\{0,+1,-1\}$ and
\begin{equation}
\label{eq:OHrelation}
{\cal H}_{\varphi,\psi}=h_{\varphi,\psi} {\cal O}_{\varphi,\psi},
\end{equation}
where
\begin{equation}
\label{eq:weight}
h_{\varphi,\psi}=\sum_{\substack{(i,j) \in \text{ non-} \\ \text{trivial loops}}}  J_{i,j} \varepsilon_{i,j}^{\varphi,\psi}+J_1{\cal L}_{\varphi,\psi}/2.
\end{equation}
Because the overlap diagram of two identical configurations contains only trivial loops, equation \eqref{eq:weight} directly implies $h_{\varphi,\varphi}=0$.

{\it Effective Hamiltonian}. Having introduced the NNVB basis $\{\ket{\psi}\}$, the overlap matrix ${\cal O}_{\varphi,\psi}$ and the matrix ${\cal H}_{\varphi,\psi}$, we can now diagonalize the Heisenberg Hamiltonian $\hat{\cal H}_H$ projected onto the singlet subspace spanned by the NNVB states. This could in principle be done in the NNVB basis directly, i.e. diagonalizing the matrix $(\hat{\cal H}_H)_{\varphi,\psi}=\langle \varphi^\star \vert \hat{\cal H}_H \vert \psi \rangle$. But as we do not want to calculate the dual basis $\{\ket{\psi^{\star}}\}$, this might not be the most practicable way. Traditionally\cite{rokhsar} one chooses the orthonormal basis $\{\ket{\psi^\perp}\}=\{\hat{\cal O}^{-1/2}\ket{\psi}\}=\{\hat{\cal O}^{1/2}\ket{\psi^\star}\}$ and diagonalizes the matrix
\begin{equation}
({\cal H}_H)_{\varphi^\perp,\psi^\perp}=\langle \varphi^\star \vert \hat{\cal O}^{1/2}\hat{\cal H}_H\hat{\cal O}^{-1/2} \vert \psi \rangle=({\cal H}_\text{eff})_{\varphi,\psi},
\end{equation}
where
\begin{equation}
\label{eq:EffectiveH}
\hat{\cal H}_\text{eff} = \hat{\cal O}^{1/2} \hat{\cal H}_H\hat{\cal O}^{-1/2} = \hat{\cal O}^{-1/2} \hat{\cal H}\hat{\cal O}^{-1/2}.
\end{equation}
Note that on the one hand-side $\hat{\cal H}_\text{eff}$ is expressed in terms of the matrices ${\cal O}_{\varphi,\psi}$ and ${\cal H}_{\varphi,\psi}$, which are both symmetric and easily calculated in terms of the loop structure formed by the NNVB states. On the other hand-side $\hat{\cal H}_\text{eff}$ arises from the Heisenberg Hamiltonian $\hat{\cal H}_H$ by a similarity transformation and therefore both operators are equivalent. However, working with $\hat{\cal H}_\text{eff}$ rather than with $\hat{\cal H}_H$ allows for using the non-orthogonal NNVB basis without any explicit knowledge of its dual basis, which is the aim of this transformation.\\
In the original scheme\cite{rokhsar,zeng-elser} the dual basis was omitted, but the definition of the effective Hamiltonian is the same in terms of $\Ov$ and $\Ham$. In order to achieve this, one had to introduce some generalized eigenvalue problem, however missing the fact that $\Ham$ and $\hat{\cal H}_H$ are indeed distinct operators (i.e. $\Ham$ is {\it not} the Heisenberg Hamiltonian). Furthermore, one had to interpret the missing $\star$ in $\sca{\varphi^{\star}}{\psi}=\delta_{\varphi,\psi}$ as emergence of so-called hardcore (or quantum) dimers\cite{rokhsar,zeng-elser,triangular-QDM,qdmkagome}. In contrast, the dual basis offers a quite natural framework to clarify the relations between the SU(2) (non-orthogonal) VB basis and the Quantum Dimer (orthogonal) basis.

\subsubsection{Operator expansion and fusions}\label{sec:expfuse}

{\it Diagrams}. In order to explicitly calculate the effective Hamiltonian $\hat{\cal H}_\text{eff}$, we need to write down all terms in $\hat{\cal O}$ and $\hat{\cal H}$. Noticing that the identity operator is given by $\hat{\cal I}=\sum_\chi\ket{\chi}\bra{\chi^\star}$, we can express the overlap operator as
\begin{subequations}\label{eq:newoverlap}
\begin{align}
\Ov = & \Ov \hat{\cal I}\nonumber\\
= & \sum_{\varphi,\psi}  \ket{\varphi}\sca{\varphi}{\psi}\bra{\psi^\star}\\
= & \sum_{\varphi,\psi} \alpha^{{\cal L}_{\varphi,\psi} - 2 {\cal N}_{\varphi,\psi}} \ket{\varphi}\bra{\psi^\star},
\end{align}
\end{subequations}
where ${\cal N}_{\varphi,\psi}$ and ${\cal L}_{\varphi,\psi}$ are number and the net length of all non-trivial loops respectively. As the amplitude ${\cal O}_{\varphi,\psi}$ (as well as $h_{\varphi,\psi}$) only depends on the non-trivial loops, it is rather instructive to replace the flipping process $\ket{\varphi}\bra{\psi^\star}$ by a diagram that visually describes the underlying resonating loop structure. For example, considering the two possible dimer coverings around a hexagon, the resulting process is
\setlength{\specdiagwidth}{0.075mm}
\begin{subequations}\label{eq:hexgraph}
\begin{align} 
\hat{\omega}_{g_1} 	&=  \kag{2_1} \label{eq:hexgraph1} \\ 
								&=  \ket{\kag{hex1}} \bra{\kag{hex2}^\star} + \ket{\kag{hex2}}\bra{\kag{hex1}^\star}, \label{eq:hexgraph2}
\end{align}
\end{subequations}
where all other loops are trivial and hence omitted in the diagrammatic representation. Note that this type of diagrams \eqref{eq:hexgraph1} actually stands for the sum of all flipping processes with the same amplitude and the same geometrical shape, i.e. all rigid motions of the diagram on the lattice are represented by one diagram. Furthermore, diagrams are symmetric, that is both processes $\ket{\varphi}\bra{\psi^\star}$ and $\ket{\psi}\bra{\varphi^\star}$ are in the same diagram. In other words we have
\begin{equation}\label{eq:graphcond}
(\og\Ov)^\dagger=\og\Ov,
\end{equation}
which we will use as determining condition for a process to be representable in a diagram.

{\it Expansion}. As $\Ov$ and $\Ham$ are closely related by \eqref{eq:OHrelation}, a simple inspection of \eqref{eq:nonorth} suggests to derive the effective Hamiltonian \eqref{eq:EffectiveH} as an expansion order by order in $\alpha$. Whereas usually\cite{rokhsar,zeng-elser} one was to work out the overlap matrix in terms of the net length ${\cal L}_{\varphi,\psi}$ of the overlap graphs, we will here consider the full exponent, i.e. additionally consider the number ${\cal N}_{\varphi,\psi}$ of loops. Therefore we will denote the order of a graph by $2\ng$ with
\begin{equation}
\label{eq:order}
\ng={\cal L}_{\varphi,\psi}/2 - {\cal N}_{\varphi,\psi},
\end{equation}
which is clearly a non-negative integer. As an example, the order of the process (\ref{eq:hexgraph}) is $2\mathrm{n}(g_1)=2(6/2-1)=4$

Then we rewrite \eqref{eq:newoverlap} as,
\begin{equation}
\label{eq:ovexpansion}
\Ov = \sg \alpha^{2\ng} \og.
\end{equation}

\setlength{\specdiagwidth}{0.09mm}
\begin{table}
\begin{equation*}
\begin{array}{||c|c|c|c|c||}\hline\hline
\text{\bf Processes} & \text{\bf Order} & \text{\bf Loop length} & \text{\bf Hexagon} & \text{\bf Degree}\\ 
\og & 2\ng & {{\cal L}_l} & \text{\bf number} & \dg\\ \hline\hline
\kag{2_1} & 4 & 6 & 1 & 1 \\
\kag{3_1} & 6 & 8 & 1 & 1 \\
\kag{4_1} & 8 & 12 & 2 & 2 \\
\kag{4_2} & 8 & 10 & 1 & 1 \\
\kag{5_1} & 10 & 14 & 2 & 2 \\
\kag{5_4} & 10 & 12 & 1 & 1 \\
\kag{5_5} & 10 & 12 & 2 & 1 \\ \hline\hline
\end{array}
\end{equation*}
\caption{\label{tab:expansionorders} Order $2\ng$ of a graph, compared to the loop length\cite{rokhsar} and the number of enclosed hexagons\cite{zeng-elser}, which were defined as order in previous works. 
Note that the different expansion schemes lead to the very same processes but these do not appear at the same orders. Nevertheless, in our procedure, the {\it leading} order
for a given loop of size ${\cal L}=2p$ ($p$ is an integer) in the ${\hat H}_{\rm eff}$ expansion is ${\cal L}-2$ (i.e. it scales linearly with the loop length).  The last column illustrates the {\it degrees} of these graphs as defined in Appendix \ref{subsec:generating}.
}
\end{table}

\setlength{\specdiagwidth}{0.045mm}
Table \ref{tab:expansionorders} compares different choices of orders and clearly shows that truncating the expansion at a given order does not produce the same terms in every case. However, our choice might seem more natural as it actually corresponds to the hierarchy of occurring amplitudes in $\Ov$. More importantly, as shown in section \ref{sec:locality}, this choice guaranties (i) the locality of $\Heff$ (i.e. non-connected processes, such as
\setlength{\specdiagwidth}{0.09mm}
\begin{equation*}
\kag{4_1} \text{ or } \kag{5_1}
\end{equation*}
appearing in $\Ham$ actually disappear in $\Heff$ {\em at all orders}) and (ii) the possibility of {\em resumming all order contributions} for a given graph.

Using \eqref{eq:oclosing} and \eqref{eq:Hamov} one can express the Hamiltonian as
\begin{subequations}\label{eq:newhamilton}
\begin{align}
\Ham = & \sum_{\varphi,\psi} \ket{\varphi} \langle \varphi \vert \hat{\cal H}_H \vert \psi \rangle \bra{\psi^\star}\\
= & \sum_{\varphi,\psi} h_{\varphi,\psi} {\cal O}_{\varphi,\psi} \ket{\varphi}\bra{\psi^\star}
\end{align}
\end{subequations}
or, similarly as \eqref{eq:ovexpansion},
\begin{equation}
\label{eq:hamexpansion}
\Ham = \sg \alpha^{2\ng} \hg \og.
\end{equation}
where $\hg=h_{\varphi,\psi}$ for the corresponding overlap graph.

The expression $\hat{\cal O}^{-1/2}\hat{\cal H}\hat{\cal O}^{-1/2}$ requires evaluating a non-integer power of $\hat{\cal O}$, which can be done using
\begin{equation}\label{eq:nonintpow}
\hat{\cal O}^\tau=\sum_{k=0}^\infty \frac{\Gamma(1+\tau)}{\Gamma(1+\tau-k)\Gamma(1+k)}(\hat{\cal O}-1)^k.
\end{equation}
However, this still requires calculating integer powers of $\hat{\cal O}$ and thus it is necessary to look at products of diagrams. Using (\ref{eq:graphcond}) for $\hat{\omega}_1 \Ov$ and $\hat{\omega}_2 \Ov$, it is easy to check that $(\hat{\omega}_1\hat{\omega}_2\Ov)^\dagger=\hat{\omega}_2\hat{\omega}_1\Ov$. Therefore a simple product $\hat{\omega}_1\hat{\omega}_2$ of diagrams does not fulfill (\ref{eq:graphcond}) and hence cannot be represented as a diagram. On the contrary, the symmetrized form
\begin{equation}
\hat{\omega}_g=\frac{1}{2}\acommute[\hat{\omega}_1]{\hat{\omega}_2}
\end{equation}
obviously verifies (\ref{eq:graphcond}) and can be represented as a graph.

{\it Fusions}. The issue of generating graphs out symmetrized products of  graphs is called \textit{fusion} of diagrams and shall be illustrated by the following fusion rules, that are valid on the kagome lattice.
\setlength{\specdiagwidth}{0.065mm}
\begin{subequations}\label{eq:fusexample}
\begin{align}
\frac{1}{2}\acommute[\kag{2_1}]{\kag{2_1}} = & \kag{2_2}+2 \kag{4_1}\label{eq:potgraph}\\
\frac{1}{2}\acommute[\kag{2_1}]{\kag{3_1}} = & \frac{1}{2}\kag{5_5} + \kag{5_1}\label{eq:fusedgraph}\\
\frac{1}{2}\acommute[\kag{2_2}]{\kag{3_3}} = & \frac{1}{2}\kag{5_14} + \kag{5_11}\label{eq:assistgraph}
\end{align}
\end{subequations}

All these rules can be obtained by direct application of processes on dimer configurations using processes definitions like \eqref{eq:hexgraph}. Let us comment  rules \eqref{eq:fusexample}.

Up to now, in the overlap matrix and the Hamiltonian we have only met flipping processes, which are called \textit{kinetic} terms. The first example (eq. \ref{eq:potgraph}) shows that the fusion of two identical kinetic diagrams always involves the emergence of \textit{potential} terms that are drawn with a yellow shape. This kind of terms arises, when a flipping process is acting twice at exactly the same position. In this case the dimers are flipped twice, which means that they are not flipped at all. However, it is important to emphasize that a potential term is not identical to the identity operator, as it is actually checked whether the plaquette is flippable or not. Indeed, applying a flipping process to a non-flippable plaquette annihilates the state. This can also be seen from the fact that contributions of potential terms can be expressed as $\ket{\varphi}\bra{\varphi^\star}$ and as an example on a hexagon we have
\setlength{\specdiagwidth}{0.075mm}
\begin{subequations}\label{eq:hexpotgraph}
\begin{align} 
\hat{\omega}_{g_2} 	&=  \kag{2_2} \label{eq:hexpotgraph1} \\ 
								&=  \ket{\kag{hex1}} \bra{\kag{hex1}^\star} + \ket{\kag{hex2}}\bra{\kag{hex2}^\star}. \label{eq:hexpotgraph2}
\end{align}
\end{subequations}

Furthermore, the fusion of diagrams always generates disconnected diagrams, where the flipping or checking processes are simply happening at distant positions (e.g. last terms on the r.h.s of equations \eqref{eq:fusexample}). Notice, that due to combinatorics the prefactor is not the same in the case when identical or different diagrams are fusing.

Equations \eqref{eq:fusedgraph} and \eqref{eq:assistgraph} illustrate, that when both kinetic or potential processes are happening close to each other (i.e. sharing at least one bond), the resulting diagrams generally look more complicated. A typical merging of two loops in one larger loop is displayed by rule \eqref{eq:fusedgraph}. Furthermore, more unusual \textit{assisted kinetic processes} where the plaquette flipping requires the presence of specific trivial loops in the neighborhood of the plaquette (see eq. \ref{eq:assistgraph}) also emerge from the fusion rules. Notice that the kinetic and potential processes are identical on trivial loops and correspond to verifying the presence of dimers at a specific position.

\setlength{\abstractdiagwidth}{0.24mm}
While the details of the fusions rules are lattice-specific, the key properties of fusions for the derivation of an effective Hamiltonian are actually quite general. In Appendix \ref{subsec:diagrammatic}, we introduce a general lattice-independent diagrammatic notation in which each connected part of a (disconnected) diagram $\mean{\gd{1}\ds\gd{1}\ldots\gd{1}}$ is generically represented as $\gd{1}$. General fusion rules (see Appendix \ref{subsec:fusion}) produce new connected terms by connecting two or more connected parts. Such terms are generically represented as $\gd{2}$ (fusion of two connected parts). For example, all the rules \eqref{eq:fusexample} lie in the generic class of rules,
\begin{equation}
\mean{\gd{1}\times\gd{1}}=\mean{\gd{2}}+\mean{\gd{11}}.
\end{equation}
Note that the notation $\mean{.}$ defined in Appendix \ref{subsec:diagrammatic}, conveniently absorbs all combinatorial prefactors. Fusing more connected parts leads to diagrams such as $\mean{\gd{3b}}$ or $\mean{\gd{3a}}$ (fusions of three connected parts). Let us remark that while both terms are connected, these two terms are inequivalent since in the latter case the first and third part do not fuse.

Provided these general fusion rules, one can work out the effective Hamiltonian $\hat{\cal H}_\text{eff}$ in a systematic way. It is important to note that $\Ov$ and $\Ham$ do not contain any potential term. They only emerge by fusing kinetic processes. On the other hand, $\Ov$ and $\Ham$ contain disconnected kinetic terms and when fusing diagrams we always generate those \textit{non-local} processes as well. Such a non-locality would be physically inconsistent in the resulting
effective Hamiltonian $\hat{\cal H}_\text{eff}=\hat{\cal O}^{-1/2} \hat{\cal H}\hat{\cal O}^{-1/2}$. However, {\em all} the non-local terms disappear while deriving $\hat{\cal H}_\text{eff}$ and thus the effective Hamiltonian is \textit{local} as the initial Heisenberg Hamiltonian. As this property, proofed in the next section, is valid at {\em all} orders in $\alpha$, this holds for the exact all-order resumed effective Hamiltonian as well as for any truncated $\hat{\cal H}_\text{eff}$.

\subsubsection{Proof of locality and resummation scheme}
\label{sec:locality}

{\it Locality}. Comparing equations \eqref{eq:ovexpansion} and \eqref{eq:hamexpansion}, the only difference between $\Ov$ and $\Ham$ is the additional weight $\hg$ for the Hamiltonian. Therefore it is convenient to define a generating function $\Z (\alpha,\beta,\mu)$ (defined as \eqref{eq:gen} in Appendix \ref{subsec:generating}), such that both $\Ov$ and $\Ham$ can be obtained from it by
\begin{subequations}
\begin{align}
\Ham & = \muzerobetaone[{\dmu[\Z]}], \\
\Ov  & = \muzerobetaone[\Z],
\end{align}
\end{subequations}
where $\beta$ and $\mu$ are some internal parameters (for details see section \ref{subsec:generating}). Therefore, $\hat{\cal H}_\text{eff}$ can be recast as
\begin{equation}
\Heff \eqmuzerobetaone \Z[-1/2] \left ( \dmu[\Z]  \right ) \Z[-1/2].
\end{equation}
After some standard operator manipulations (see section \ref{subsec:logZ}), one can show, that the last equation is equivalent to
\begin{equation}
\label{eq:Heffcommutators}
\Heff \eqmuzerobetaone \sum_{p=0}^{\infty} \frac{1}{2^{2p}} \frac{\left ( \halfcommute[\logZ] \right )^{2p}}{\left ( 2 p + 1 \right )!} \dlogZ.
\end{equation}
involving only iterated commutators of $\logZ$ on $\dlogZ$.

Now, supposing that $\logZ$ only contains connected terms, it is clear that this cannot be different in $\dlogZ$. Moreover, the commutator of two connected diagrams cannot be disconnected: (i) either the two diagrams act on disconnected plaquettes and hence commute, (ii) or they lie on neighboring plaquettes (i.e. sharing at least one bond) and they fuse to a single connected diagram. This can also be shown easily using $[A,[A,B]]=\{B,\{A,A\}\}-\{A,\{A,B\}\}$ and the diagrammatic scheme in appendix (\ref{subsec:diagrammatic}).

It turns out, that $\logZ$ indeed only contains connected terms, which is an important result both from a conceptual and aesthetic points of view: it shows that $\Heff$ is local and therefore provides firm grounds to the consistency of the method. Conversely, producing an effective Hamiltonian with non-local terms would put the whole scheme into question. 

The quite technical demonstration of this point is split in several Appendices (\ref{subsec:cumulants} to \ref{sec:linkedcluster}). Let us sketch the arguments. One can establish a linked cluster theorem, i.e. express the logarithm of the generating function as sum of cumulants (see Appendix \ref{subsec:cumulants}). One can then formulate some lattice independent diagrammatic notation (Appendix \ref{subsec:diagrammatic}) and work out general fusion rules (Appendix \ref{subsec:fusion}). Using these rules, we establish equation \eqref{eq:recursion4} showing that every cumulant can be reexpressed as a combination of lower order cumulants.
This allows finally to show by mathematical induction that cumulants represent connected processes and therefore $\logZ$ does not contain non-connected terms (appendix \ref{sec:linkedcluster}). As we will see, this generating function formalism is not only useful to establish the effective Hamiltonian locality but also provides a practical framework to resum the 
series in $\alpha$ giving the amplitude of dominant terms in $\Heff$.

{\it Fully connected diagrams}. A special class of connected diagrams, called \textit{fully connected} diagrams is produced by complete fusion of all parts (i.e every diagram is connected to every other diagram by sharing at least one bond). In the lattice-independent diagrammatic representation, we denote such diagrams as
\setlength{\abstractdiagwidth}{0.35mm}
\begin{equation*}
\bgd{NFullyConnected}
\end{equation*}
These diagrams are very important for two reasons:
\begin{itemize}
\item This class of diagrams includes the most compact terms (i.e. involving short loops) of the effective Hamiltonian that are obtained by fusing identically shaped diagrams exactly at the same place (e.g. the first term in the r.h.s of eq. \eqref{eq:potgraph}). These processes are also the dominant terms in the $\alpha$-expansion of $\Heff$. Indeed, the leading order for a length-${\cal L}$ connected process in $\Heff=\hat{\cal O}^{-1/2} \hat{\cal H}\hat{\cal O}^{-1/2}$ is $\alpha^{{\cal L}-2}$.
\item Their exact weight in $\logZ$ is particularly simple to obtain, as explained in appendix (\ref{sec:fullyconnected}) and, 
\setlength{\abstractdiagwidth}{0.24mm}
\begin{equation}
\label{eq:maximal}
\begin{split}
\logZ&=-\sum_{m=1}^{\infty} \frac{(-1)^m}{m!} \beta^m \mean{{\overbrace{\left(\bgd{NFullyConnected}\right)}^{m}}_c}\\
&\quad + {\text{less connected terms}}.
\end{split}
\end{equation}
\end{itemize}

\setlength{\specdiagwidth}{0.09mm}
{\it Elementary diagrams \& Resummation}. We define the {\it elementary diagrams} as the set of processes (kinetic or potential) that cannot be cut into two sub-processes. For example, 
\begin{equation}
\label{eq:elemexample}
\kag{2_1} \text{ and } \kag{3_1}
\end{equation}
are elementary while
\begin{equation}
\label{eq:nonelemexample}
\kag{5_5}
\end{equation}
is not since it can be cut into the two parts \eqref{eq:elemexample}.
Obviously, all elementary graphs appear as fully connected graphs in $\logZ$: for example $\hat{\omega}_{g_2}$ given by \eqref{eq:hexpotgraph} is a full connection of $\hat{\omega}_{g_1}$, see \eqref{eq:hexgraph}, with itself. \setlength{\specdiagwidth}{0.075mm}Notice that there exist fully connected terms that are not elementary (e.g. \eqref{eq:nonelemexample} is a full connection of \eqref{eq:elemexample}, but is not elementary). 

\setlength{\abstractdiagwidth}{0.35mm}
Generic elementary kinetic and potential diagrams will be represented respectively as,
\begin{equation}
\label{eq:genericelem}
\bgd{ElemKin} \;\;\;\;\text{ and }\;\;\;\; \bgd{ElemPot}.
\end{equation}
Fusion rules of elementary diagrams, of which \eqref{eq:potgraph} is a typical example, can be compactly written as
\begin{subequations}\label{eq:elemrules}
\begin{align}
\frac{1}{2}\acommute[\bgd{ElemKin}]{\bgd{ElemKin}} = & \; \bgd{ElemPot} + 2 \; \bgd{ElemKinKin} \label{eq:elemruleskin}\\
\frac{1}{2}\acommute[\bgd{ElemPot}]{\bgd{ElemPot}} = & \; \bgd{ElemPot} + 2 \; \bgd{ElemPotPot} \label{eq:elemrulespot}\\
\frac{1}{2}\acommute[\bgd{ElemKin}]{\bgd{ElemPot}} = & \; \bgd{ElemKin} +  \bgd{ElemKinPot} \label{eq:elemrulesmixted}
\end{align}
\end{subequations}

In order to compute the full weight of elementary diagrams in $\logZ$, we need to track all occurrences of these terms in \eqref{eq:maximal}. High order contractions of these processes are typically given by iterating \eqref{eq:elemrules}. But obviously, the last (disconnected) terms of \eqref{eq:elemrules} cannot produce fully connected terms by further iterations. As a consequence, the evaluation of $\logZ$ only requires the simple reduced rule,
\begin{equation}
\label{eq:reduced}
\bgd{ElemKin}^n =
	\begin{cases}
	\bgd{ElemKin}& \text{for odd } n,\\
	& \\
	\bgd{ElemPot}& \text{for even } n.
\end{cases}
\end{equation}

Since a length-${\cal L}$ elementary process with weight $h$ in the expansion of $\Ham$ occurs at order $\alpha^{{\cal L}-2}$, its contribution to the generating function is,
\begin{equation}
\label{eq:elemZ}
\Z = \ldots + \beta \alpha^{{\cal L}-2} e^{\mu h} \; \bgd{ElemKin} + \ldots
\end{equation}
Using \eqref{eq:maximal}, the relevant contribution to $\logZ$ is,
\begin{equation}
\label{eq:elemlogZ}
\logZ = \ldots + \sum_{m \geq 1} \beta^m \frac{(-1)^m}{m} \alpha^{m({\cal L}-2)} e^{m \mu h} \; \bgd{ElemKin}^m + \ldots
\end{equation}
which is easily evaluated, using \eqref{eq:reduced}, as
\begin{align}
\label{eq:elemlogZexplicit}
\logZ = \ldots &+ \frac{1}{2} \ln \frac{1+\beta \alpha^{{\cal L}-2} e^{\mu h}}{1-\beta \alpha^{{\cal L}-2} e^{\mu h}}\;\bgd{ElemKin}\\ \nonumber
&+ \frac{1}{2} \ln \left ( 1-\beta^2 \alpha^{2({\cal L}-2)} e^{2 \mu h} \right )\;\bgd{ElemPot}+ \ldots
\end{align}

Evaluating $\Heff$ is done using \eqref{eq:Heffcommutators} which requires in turn to compute $\dlogZ$ and all its iterated commutators with $\logZ$. But since, by definition, elementary diagrams cannot be produced by fusing smaller diagrams, the only relevant terms to the amplitude of \eqref{eq:genericelem} in $\Heff$ produced by orders $p>0$ in \eqref{eq:Heffcommutators}
are (i) commutators of an elementary diagram with itself at the same place or (ii) commutators of two elementary diagrams acting on disconnected places. Obviously such contributions vanish identically. As a direct consequence, only order $p=0$ of \eqref{eq:Heffcommutators} contribute to the amplitude of elementary diagrams in $\Heff$. In other words, for any elementary process
\begin{equation}
\label{eq:Heffcommutatorselem}
\Heff \eqmuzerobetaone \dlogZ.
\end{equation}
It is then straightforward, using \eqref{eq:elemlogZexplicit}, to evaluate the contribution of elementary diagrams to $\Heff$ :
\begin{align}
\label{eq:elemHeff}
\Heff = \ldots &+ h \frac{\alpha^{{\cal L}-2}}{1-\alpha^{2({\cal L}-2)}} \;\bgd{ElemKin}\\ \nonumber
&- h \frac{\alpha^{2({\cal L}-2})}{1-\alpha^{2({\cal L}-2)}}\;\bgd{ElemPot}+ \ldots
\end{align}
where ${\cal L}$ is the length of the elementary (kinetic) diagram and $h$ its ``bare'' energy as appearing in the $\Ham$ expansion \eqref{eq:hamexpansion}.

\section{Generalized Quantum dimer model for the kagome antiferromagnet}
\label{sec:kagome}

In this section we will apply our scheme to the kagome lattice and consider the nearest neighbor coupling $J_1=J>0$ only.
First, we make a brief review of the current understanding of this model.

\subsection{Current status}

The properties of the QHAF on the Kagome lattice have been actively explored 
for the last two decades by analytical and numerical techniques.  Although a number of important results have 
been obtained, the nature of the groundstate is still a mystery.
Lanczos ED of small
clusters~\cite{lecheminant,sindzingre-99} have clearly revealed an exponentially large number (w.r.t. system size) 
of singlets below the first triplet excitation (spin gap), a fact reminiscent of the NNVB basis~\cite{mila,kagomeNNVB}. The accessible cluster 
sizes remain however too small
to definitely conclude whether the spin gap survives in the thermodynamic limit~\cite{sindzingre-lhuillier}. 
Recent DMRG studies~\cite{dmrg} seem however to support it.

A number of studies have also been devoted more precisely to the nature of the groundstate itself.
For example, an early ED analysis of the four-spin correlations (i.e. dimer-dimer) first pointed towards a {\it short-range} dimer 
liquid phase~\cite{leung}. 
Alternatively, translation-symmetry breaking VBC, two-dimensional analogs of the Majumdar-Gosh~\cite{mg} 
dimerized chain, have been proposed on the basis of various other approaches like large-N SU(N) techniques~\cite{vbc,marston} and 
mappings to low-energy effective hamiltonians within the singlet subspace~\cite{maleyev,senthil,auerbach}.
Recent series expansions around the dimer limit~\cite{singh} showed that a 36-site unit cell (i.e. a $2\sqrt{3}\times 2\sqrt{3}$ supercell
of up-triangles) VBC would be preferred in agreement with Refs.~\onlinecite{marston,senthil}.
However, the interpretation of the ED low-energy singlet spectrum~\cite{misguich-sindzingre} on the 36-site periodic cluster remains problematic 
not showing the expected quasi-degenerate VBC groundstates. Therefore, it has been proposed~\cite{sindzingre-lhuillier} that the 
Heisenberg model might be within (or in the close vicinity of) a spin liquid phase such as the algebraic spin liquid~\cite{lee-et-al}. 
We now turn to the application of our procedure which offers a new versatile scheme to investigate non-magnetic singlet 
phases as VBC.

\subsection{Expansion}\label{sec:Heffkagomeexp}

{\it Leading orders}. Table \ref{tab:expansionHO} shows all the contributing terms in $\Ov$ and $\Ham$ up to order $2\ng=10$. Of course, as ${\cal O}_{\varphi,\varphi}=1$ and ${\cal H}_{\varphi,\varphi}=0$, the identity process $\1$, which only contains trivial loops, appears with order $2\ng=0$ in $\Ov$ but does not contribute to $\Ham$. Therefore we define $\Ov=\1+\hat{\cal X}$, where $\hat{\cal X}$ is a short hand notation for all processes in $\Ov$, except the identity.

\setlength{\specdiagwidth}{0.075mm}
\begin{table}
\begin{equation*}
\begin{array}{||c|c|c||c|c|c||}\hline\hline
\text{\bf Processes} & \hat{\cal O} & \hat{\cal H}/J & \text{\bf Processes} & \hat{\cal O} & \hat{\cal H}/J \\ \hline\hline
\kag{2_1} & \alpha ^4 & -3  \alpha ^4 &  \kag{5_1} & \alpha ^{10} & -5 \alpha ^{10} \\
\kag{3_1} & \alpha ^6 & -2  \alpha ^6 &  \kag{5_2} & \alpha ^{10} & -5 \alpha ^{10}  \\
\kag{3_2} & \alpha ^6 & -2  \alpha ^6 &  \kag{5_3} & \alpha ^{10} & -5 \alpha ^{10}  \\
\kag{3_3} & \alpha ^6 & -2  \alpha ^6 &  \kag{5_4} & \alpha ^{10} & 0 \\
\kag{4_1} & \alpha ^8 & -6  \alpha ^8 &  \kag{5_5} & \alpha ^{10} & -4 \alpha ^{10}  \\
\kag{4_2} & \alpha ^8 & -\alpha ^8 &  \kag{5_6} & \alpha ^{10} & -4 \alpha ^{10}  \\
\kag{4_3} & \alpha ^8 & -\alpha ^8 &  \kag{5_7} & \alpha ^{10} & -4 \alpha ^{10}  \\
\kag{4_4} & \alpha ^8 & -\alpha ^8 &  \kag{5_8} & \alpha ^{10} & -4 \alpha ^{10}  \\ \hline\hline
\end{array}
\end{equation*}
\caption{\label{tab:expansionHO} Expansion of $\hat{\cal X}=\hat{\cal O}-\1$ and $\hat{\cal H}$ up to order $2\ng=10$.}
\end{table}

Using equation (\ref{eq:nonintpow}) the effective Hamiltonian $\Heff = \Ov[-1/2] \Ham \Ov[-1/2]$ can be written as
\begin{align}\label{eq:expbyhand}
\Heff = & \Ham - \frac{1}{2}\acommute[\hat{\cal X}]{\Ham} \\
+ & \frac{1}{8}\acommute[\hat{\cal X}]{\acommute[\hat{\cal X}]{\Ham}} + \frac{1}{8}\acommute[\Ham]{\acommute[\hat{\cal X}]{\hat{\cal X}}} \nonumber \\
+ & \text{ terms with at least 3 anticommutators.} \nonumber
\end{align}
Thus we can immediately see, that all elementary kinetic processes have, at the lowest order, the same weight in $\Ham$ and in $\Heff$ (compare table \ref{tab:expansionHO} with \ref{tab:expansionHeff}). Additionally considering the second term in (\ref{eq:expbyhand}) and with the help of fusion rule (\ref{eq:potgraph}), one can confirm the amplitude of
\setlength{\specdiagwidth}{0.09mm}
\begin{equation*}
\kag{2_2}
\end{equation*}
to be $-\alpha ^4(-3 J \alpha ^4)=3 J \alpha ^8$ (see Table \ref{tab:expansionHeff}). Similarly, it is easy to see that the amplitude of
\begin{equation*}
\kag{4_1}
\end{equation*}
cancels out to $-2\alpha ^4(-3 J \alpha ^4)+(-6 J \alpha ^8)=0$ which is a direct verification of the general property proved in Section \ref{sec:locality} stating that disconnected graphs do not contribute to the effective Hamiltonian at any order. The leading orders of all the diagrams presented in Table \ref{tab:expansionHeff} can be obtained conveniently using a direct evaluation based on \eqref{eq:expbyhand}. 

\setlength{\specdiagwidth}{0.075mm}
\begin{table}
\begin{equation*}
\begin{array}{||c|c|c||c|c|c||}\hline\hline
\multirow{2}{*}{\text{\bf Processes}} & \multicolumn{2}{c||}{{\Heff} / J}  & \multirow{2}{*}{\text{\bf Processes}} & \multicolumn{2}{c||}{{\Heff} / J} \\ \cline{2-3} \cline{5-6}
 & \text{\bf LO} & \infty &  & \text{\bf LO} & \infty \\  \hline\hline
\kag{2_1} & -3  \alpha ^4 & -\frac{3  \alpha ^4}{1-\alpha ^8} & \kag{4_2} & - \alpha ^8 & -\frac{\alpha ^8}{1-\alpha ^{16}} \\
\kag{2_2} & 3  \alpha ^8 & \frac{3  \alpha ^8}{1-\alpha ^8} &  \kag{4_3} & - \alpha ^8 & -\frac{\alpha ^8}{1-\alpha ^{16}} \\
\kag{3_1} & -2  \alpha ^6 & -\frac{2  \alpha ^6}{1-\alpha ^{12}} & \kag{4_4} & - \alpha ^8 & -\frac{\alpha ^8}{1-\alpha ^{16}} \\
\kag{3_2} & -2  \alpha ^6 & -\frac{2  \alpha ^6}{1-\alpha ^{12}} & \kag{4_6} &  \alpha ^{16} & \frac{\alpha ^{16}}{1-\alpha ^{16}} \\
\kag{3_3} & -2  \alpha ^6 & -\frac{2  \alpha ^6}{1-\alpha ^{12}} & \kag{4_7} &  \alpha ^{16} & \frac{\alpha ^{16}}{1-\alpha ^{16}} \\
\kag{3_4} & 2  \alpha ^{12} & \frac{2  \alpha ^{12}}{1-\alpha ^{12}} & \kag{4_8} &  \alpha ^{16} & \frac{\alpha ^{16}}{1-\alpha ^{16}} \\
\kag{3_5} & 2  \alpha ^{12} & \frac{2  \alpha ^{12}}{1-\alpha ^{12}} &  \kag{5_4} & 0 & 0 \\
\kag{3_6} & 2  \alpha ^{12} & \frac{2  \alpha ^{12}}{1-\alpha ^{12}} & \kag{5_16} & 0 & 0 \\  \hline\hline
\end{array}
\end{equation*}
\caption{\label{tab:expansionHeff} Elementary processes in ${\cal H}_{\text{eff}}$ at Leading Order (LO) and fully resummed series ($\infty$).}
\end{table}

{\it Higher orders}. Diagrams with larger leading orders (typically enclosing several hexagons) as well has 
further renormalizing corrections to low leading order diagrams are more problematic to obtain using simple arguments. Nevertheless, the calculation can be systematically extended to significantly higher orders using the expansion scheme presented in Section \ref{sec:expfuse}. Up to the chosen order, this requires (i) the enumeration of all terms appearing in $\Ham$, (ii) a careful enumeration of fusion rules that proliferate as the order increases, (iii) the expansion of $\Ov[-1/2]$ and (iv) the evaluation of $\Ov[-1/2] \Ham \Ov[-1/2]$. Note that
the two last steps explicitly require using the fusion rules obtained in step (ii). The details and results of such a procedure up to order $\alpha^{14}$ are too lengthy to be presented in this article and are therefore provided as supplementary material at (see at the end of the article) in which we include all the relevant fusion rules up to order $\alpha^{14}$ as well as extensions of Tables \ref{tab:expansionHO} and \ref{tab:expansionHeff}.\\

\subsection{Resummation}\label{sec:Heffkagomediscuss}

A simple inspection of Table \ref{tab:expansionHeff} reveals that all the leading processes in the effective Hamiltonian are elementary diagrams in the sense defined in Section \ref{sec:locality}. Indeed, none of the terms enclosing only one hexagon
can be split in sub-processes. This important remark shows that the resummation scheme of elementary diagrams presented above applies directly. In particular the equation \eqref{eq:elemHeff} immediately leads to the resummed amplitudes of the effective Hamiltonian presented in Table \ref{tab:expansionHeff}. The explicit form of $\Heff$ is obtained by setting $\alpha=i/\sqrt{2}$ which leads to
\begin{widetext}
\setlength{\specdiagwidth}{0.09mm}
\begin{align}\label{eq:Hefffull}
\Heff / J = &- \frac{4}{5}\kag{2_1} + \frac{1}{5}\kag{2_2} + \frac{16}{63}\left( \kag{3_1} + \kag{3_2} + \kag{3_3} \right) \\
						&+ \frac{2}{63} \left( \kag{3_4} + \kag{3_5} + \kag{3_6} \right) - \frac{16}{255} \left( \kag{4_2} + \kag{4_3} + \kag{4_4} \right) \nonumber \\
						&+ \frac{1}{255} \left( \kag{4_6} + \kag{4_7} + \kag{4_8} \right) + 0 \left( \kag{5_4} + \kag{5_16} \right)  \nonumber
\end{align}
\end{widetext}

\subsection{Discussion}\label{sec:Heffkagomeresum}
{\it General remarks}. Note that the kinetic part of this Hamiltonian~\cite{note-sign1,note-sign2} is quite close to the one originally proposed by Zeng \& Elser\cite{zeng-elser} with only small differences in the magnitudes of the processes, differences introduced by 
our infinite order resummation scheme. This provides strong evidence that the expansion indeed converges rapidly.
However, very importantly, our Hamiltonian includes also diagonal terms which turn out to play a major role but which were not included in equation (7) of Ref.~\onlinecite{zeng-elser} based on a different expansion scheme~\cite{note-zeng-elser}. Indeed, the low-energy gap presented on figure 3 in Ref.~\onlinecite{zeng-elser}, whose magnitude is approximately $36 \times 0.0055 J \approx J/5$, splits two sectors with  $2$ and $1$ flippable hexagon(s) respectively (see figure 4 in the same reference). Yet, this gap value 
of $J/5$ is precisely the amplitude of the potential term, disadvantaging flippable hexagons, obtained in the effective Hamiltonian \eqref{eq:Hefffull}. This strongly suggests that this gap, known to be absent from the exact spectrum of the Heisenberg Hamiltonian\cite{waldtmann}, is an artifact produced by truncating the expansion. As already mentioned (see Section \ref{sec:expfuse} and Table \ref{tab:expansionorders}) the traditional expansion scheme\cite{rokhsar,zeng-elser} as a tendency to push away to higher orders in $\alpha$ the emergence of terms in the effective Hamiltonian. On the contrary, including such a potential term is likely to close this gap
which brings Hamiltonian \eqref{eq:Hefffull} closer to the actual low-energy phenomenology of the kagome antiferromagnet.

Another important remark is that the amplitude of the kinetic (and potential) pinwheel process, 
\setlength{\specdiagwidth}{0.09mm}
\begin{equation*}
\kag{5_4}
\end{equation*}
denoted $J_{12}$ below, exactly vanishes at all orders. As discussed below including a finite $J_{12}$ in the model 
lifts a very special degeneracy of the groundstate manifold.

{\it Large scale numerical computation}. In contrast to the case of the frustrated square lattice~\cite{j1j2j3dimer}, for the kagome lattice we obtain an effective model whose leading coefficients have alternating signs 
precluding any stochastic approach. However, Lanczos exact diagonalizations can be preformed on relatively large clusters~\cite{qdmkagome} due to the very constrained nature 
of the dimer basis that greatly limits the number of states ($2^{\frac{N}{3}+1}$ compared to $2^N$ for SU(2) spin-1/2 models).  
Furthermore, group theory techniques can be applied to block-diagonalize the Hamiltonian matrix in each of its irreducible representations (IRREP)
hence further reducing the practical number of degrees
of freedom. For example,  for the most interesting clusters with $N=3n^2$ or $9m^2$ sites (which possess all relevant space group symmetries of the 
infinite lattice) such as the 36-, 48-, 108- and 144-sites
clusters, the increasing Hilbert space sizes of their smallest (largest) IRREP are roughly of the order of 15 (170), 70 (2$\times 10^3$), 80$\times 10^{6}$ (950$\times 10^{6}$) , 200$\times 10^{9}$ (3$\times 10^{12}$) respectively. Hence, current supercomputers enable to tackle the 108-sites cluster while the larger 144-site cluster might be reachable within a few decades.

{\it Model properties}. The numerical results for Hamiltonian \eqref{eq:Hefffull} presented in Ref.~\onlinecite{qdmkagome} (corresponding in fact to an earlier extremely close 14-th order estimation of the
model) have been summarized in the phase diagram of Fig.~(\ref{fig:phases}).  Here, we have introduced two extra parameters; (i) $\lambda$ to interpolate (linearly) between 
the ``Heisenberg point''  \eqref{eq:Hefffull} at $\lambda=1$ and the ``RK-point'' of Ref.~\onlinecite{misguich-RK} (denoted $\hat{{\cal H}}_{RK}$ below) at $\lambda=0$. (ii) A finite pinwheel amplitude $J_{12}$. This double interpolation can be summarized in the two-parameter Hamiltonian
\begin{align}
\label{eq:hsim}
\hat{{\cal H}}_{\text{interp.}} (\gamma,J_{12}) 	
&= \gamma \Heff + (1-\gamma) \hat{{\cal H}}_{RK} \\
&+ \left ( J_{12} +\frac{1}{4} (\gamma-1) \right ) \kag{5_4},\nonumber
\end{align}
where $J=1$ and the RK-Hamiltonian has been defined in Ref.~\onlinecite{misguich-RK} with $\Gamma=-1/4$~\cite{note-sign1}.

\begin{figure}
\begin{center}
\includegraphics[width=0.95\linewidth]{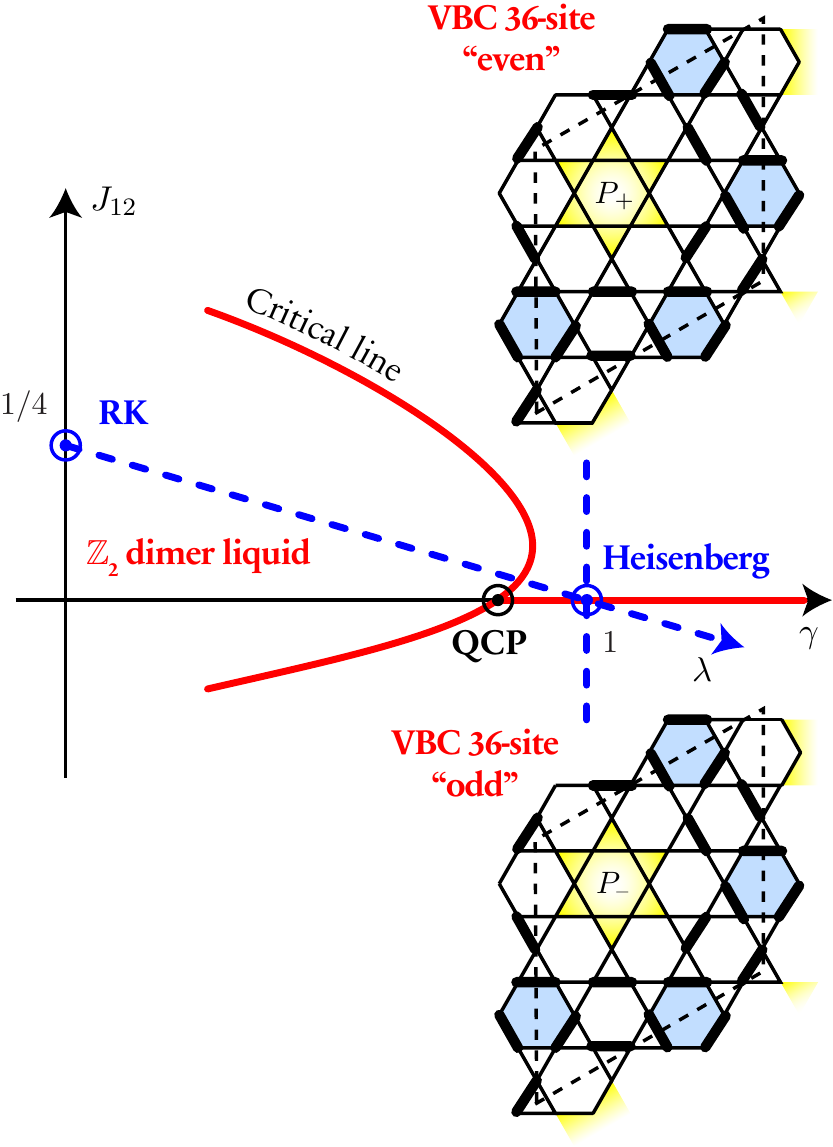}
\end{center}
\caption{\label{fig:phases} (Color online). Semi-quantitative phase diagram of the extended GQDM for the Kagome lattice as a function of the
pinwheel resonance amplitude $J_{12}$ and  the parameter $\gamma$ (see Eq.~\eqref{eq:hsim}). The two dashed blue lines represent (i) an interpolation $\lambda \Heff + (1-\lambda) \hat{{\cal H}}_{RK}$ between
the RK model and the effective model for the Heisenberg quantum antiferromagnet and (ii) a cut at $\gamma=1$ along $J_{12}$,
and correspond to the simulations performed in reference~\onlinecite{qdmkagome}. Red lines are qualitative phase transitions. Note that for $\gamma > 0.9$ a finite $J_{12}$ lifts the degeneracy between two VBC's with identical 36-site unit cells
but opposite parities $P_{\pm}$ of their resonating pinwheels (in yellow). }
\end{figure}

In agreement with series expansions~\cite{singh} and earlier work~\cite{marston,senthil} on the QHAF, the GS of our corresponding effective model (\ref{eq:Hefffull}) is found to be a VBC with a $2\sqrt{3}\times2\sqrt{3}$ supercell of up-triangles (carrying 36 sites) whose underlying dimerization pattern corresponds to an honeycomb-lattice arrangement of the resonating ``perfect'' hexagons (colored in blue in Fig.~(\ref{fig:phases})). Because $J_{12}=0$ the Heisenberg effective model (\ref{eq:Hefffull})
is also characterized by a degeneracy of the even and odd resonating pinwheels of the motive (colored in yellow in Fig.~(\ref{fig:phases})).
As pointed out in Ref.~\onlinecite{singh} this leads to an extra Ising-like degeneracy. Remarkably, the numerical results show also that the
``Heisenberg point'' lies very close to the critical line which separate the VBC phase from an extended dimer liquid phase similar to 
the one at the ``RK point''.  Interestingly, a general field-theoretic framework~\cite{sachdev-QCP} based on a double Chern-Simons theory
correctly describes such a Quantum Critical Point: one considers the spectrum of visons (i.e. topological defects) in the dimer (so-called $\mathbb{Z}_2$) liquid phase,
and studies how they condense -- the condensation of visons leads VBC order. Our approach applied to the Kagome lattice as well as other
numerical studies of a generic QDM on the triangular lattice~\cite{triangular-condensation} strongly supports such a scenario. Even though the $\mathbb{Z}_2$ phase has no broken symmetry, it does have a topological order. In principle, the topological
order can co-exist with the VBC, and so their vanishing at a common critical point without fine-tuning, can be considered as
a non-LGW (Landau-Ginzburg-Wilson) transition~\cite{non-LGW}.

\section{Concluding remarks}\label{sec:conclusion}

In summary, a systematic non-perturbative method has been developed to describe quantitatively the low-energy physics of frustrated QHAF. Provided that the latter 
is governed by fluctuations of (short-range) singlets (which is believed to happen in many cases), this method is fairly general and can be applied in principle to any lattice geometry. 
The low-energy effective model takes the form of a Generalized Quantum Dimer Model which is proven to be local, a physical requirement. 
Complete practical formalism is presented as well as other important general results.
It is also shown that the expansion scheme can be pursued up to high orders and the most relevant terms can be resummed.

As a practical implementation of the method, we have considered the much debated Kagome QHAF where a fully resummed parameter-free GQDM 
can be obtained. Although, the resulting model bears a sign problem (e.g. could not be simulated by QMC techniques), ED results up to 
clusters with 108 sites can be performed~\cite{qdmkagome} showing strong evidence in favor of a large supercell VBC. 
Also, it turns out that, in some extended parameter space, the effective model of the Kagome antiferromagnet lies in the close vicinity
of a critical line towards a topological quantum dimer liquid phase, somehow clarifying the low-energy puzzle of the original spin model. 
Let us recall that a similar approach has also been applied to the frustrated square lattice~\cite{j1j2j3dimer}.
This shows that effective GQDM can efficiently describe the most frustrated quantum magnets and greatly help to understand their properties on larger (smaller) length (energy) scales compared e.g. to standard ED techniques. It could be used also in 3 dimensions 
to tackle the quantum antiferromagnet e.g. on the hyper-kagome lattice. 

A number of Hamiltonian extensions could easily be included in the following approach like (i) other SU(2)-invariant terms as multiple exchange or
longer-range exchange interactions, (ii) doping with static or mobile holes~\cite{holons} and (iii) the inclusion of spinons~\cite{spinons} and triplets.
In the case of the Kagome lattice, it is important whether small non-SU(2) terms like Dzyaloshinski-Moriya interaction could also be inserted 
in a realistic way. All these issues are left for future studies.

\setlength{\abstractdiagwidth}{0.23mm}
\section{Appendices}
\label{sec:appendices}
\subsection{Dimers, overlaps and sign conventions}
\label{subsec:signs}

In this appendix we recall some of the most important basic properties of VB states and two
orientation conventions that are suitable for the derivation scheme proposed in this article, respectively for bipartite and non-bipartite lattices.

{\it Overlaps}. Let us consider a Valence Bond (VB) wavefunction
\begin{equation}
\label{eq:VB}
\ket{\varphi} = \prod_{k=1}^{N/2} [i_k,j_k],
\end{equation}
where
\begin{equation}
\label{eq:SU2Dimer}
[i,j]=\frac{1}{\sqrt{2}} \left ( \ket{\uparrow_{i} \downarrow_{j}} - \ket{\downarrow_{i} \uparrow_{j}} \right )
\end{equation}
is the SU(2) dimer wavefunction on sites $i$ and $j$. Two arbitrary VB states $\ket{\varphi}$ and $\ket{\psi}$ are non-orthogonal and
\begin{equation}
\label{eq:overlap}
	\sca{\varphi}{\psi} = \eta \left ( \varphi,\psi \right )  2^{n_l \left ( \varphi,\psi \right )-N/2}
\end{equation}
where $n_l$ is the number of loops of the overlap diagram (see figure \ref{fig:overlap}) and $\eta \left ( \varphi,\psi \right )$ is a sign that depends on the chosen convention for orientating SU(2) dimers.

\begin{figure}
\begin{center}
\includegraphics[width=\linewidth]{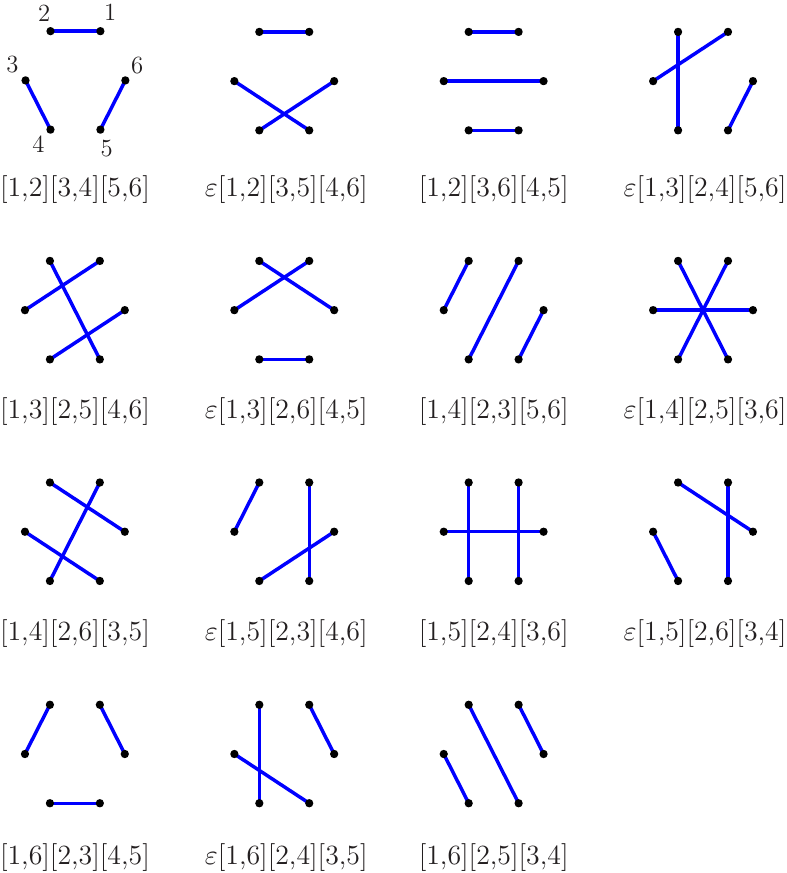}
\end{center}
\caption{\label{fig:signs} Explicit bosonic ($\varepsilon=+1$) and fermionic ($\varepsilon=-1$) conventions for the 15 VB states on 6 sites.}
\end{figure}

The question of orientating dimers in order to fix the form of $\eta_{\varphi,\psi}$ is determined in turn by two parameters: (i) the nature (bipartite or non-bipartite) of the lattice on which the dimers are constructed and (ii) the constraints one puts on the type of dimers considered (arbitrary long range dimers, arbitrary long range dimers with a bipartite constraint, nearest neighbor dimers,\ldots). In the perspective of performing an expansion of the overlap matrix as powers of a small parameter it is essential to choose conventions where the sign in the power law of equation (\ref{eq:overlap}) can be systematically absorbed: 
\begin{equation}
\label{eq:overlap2}
	\sca{\varphi}{\psi} = \alpha^{N-2 n_l \left ( \varphi,\psi \right )}
\end{equation}
The only two possible conventions denoted $b$ (bosonic) and $f$ (fermionic) for which
\begin{align}
\label{eq:bf}
\eta_{f}^{b} \left ( \varphi,\psi \right )=(\pm 1)^{n_l(\varphi,\psi)-N/2},
\end{align}
correspond to  $\alpha_b = 1 / \sqrt{2}$ and $\alpha_f = i /\sqrt{2}$. These two representations, widely discussed in the literature (see for example Ref.~\onlinecite{read}), originate from the fact that dimer wavefunctions can be constructed out of a vacuum by applying bosonic or fermionic operators resulting in different statistic and hence sign conventions. In the following paragraph we willfully choose to present these two conventions without any direct reference to bosonic or fermionic operators, emphasizing the fact it can be directly understood as a conventional orientation of dimers, as made explicit in Figure~\ref{fig:signs} for 6 sites.

{\it Bosonic convention on a bipartite lattice}. In this convention, the set of $N$ sites is divided into two $N/2$-site subsets $\cal A $ and $\cal B$, and dimers $[i,j]$ are oriented such as $i\in {\cal A}$ and $j \in {\cal B}$. As a direct consequence of this orientation, arrows are diverging from all $\cal A$ sites and converging to all $\cal B$ sites in the overlap diagram. The contribution of each loop is thus $+2$ and $\alpha=1/\sqrt{2}$.

Note that this convention can also be used on non-bipartite lattices, provided an arbitrary bond orientation is specified. However, this generally produces a sign in the overlaps that cannot absorbed such as in \eqref{eq:overlap2}.

{\it Fermionic convention}.
Choosing an arbitrary numbering of sites, we start from a reference configuration,
\begin{equation}
\label{eq:ref}
	\ket{\varphi_0} = [1,2] [3,4] \ldots [N-1,N].
\end{equation}
Here on a non-bipartite lattice, in the bosonic representation of the dimers
we need to specify the sign of each dimer following some prescription.
Therefore, by convention, we write the dimer wave function (\ref{eq:SU2Dimer}) such as $i_k < j_k$. 
In the expression
\begin{equation}
\label{eq:ordre}
{\prod_{k=1}^{N/2}}^{*} [i_k,j_k],
\end{equation}
the symbol $*$ means that $i_k < j_k$.
Any state (\ref{eq:ordre}) can be obtained from $\ket{\varphi_0}$ by applying a {\it unique} permutation  $\pi$ acting on site numbers
\begin{equation}
\label{eq:perm}
	\pi = \left (
\begin{array}{ccccc}
	1 	& 2 	& \ldots & N-1 		& N 		\\
	i_1 & j_1 & \ldots & i_{N/2} 	& j_{N/2}	\\
\end{array}
		\right ).
\end{equation}
Then we define 
\begin{equation}
\label{eq:orientation}
	\ket{\varphi} = \sigma(\pi) {\prod_{k=1}^{N/2}}^{*} [i_k,j_k],
\end{equation}
where $\sigma(\pi)$ is the signature of the permutation $\pi$ :
\begin{equation}
\label{eq:signature}
	\sigma(\pi)= \sum_{i<j} \frac{\pi(i) - \pi(j)}{i-j}.
\end{equation}

Note that (\ref{eq:orientation}) is in fact independent of the prescription to order the individual
dimers but depends only on the ordering of the sites in the reference state (\ref{eq:ref}). 
An explicit example of this convention (as well as the bosonic convention) is given on figure \ref{fig:overlap}. Let us remark that if $\vert \varphi \rangle$ and $\vert \psi \rangle$ denote two states deduced from the reference state (\ref{eq:ref}) by applying the permutations $\pi_\varphi$ and $\pi_\psi$ then their relative sign according to (\ref{eq:orientation}) is $\sigma(\pi)$ where $\pi=\pi_\varphi \pi_\psi$.

We are going to show that convention (\ref{eq:orientation}) indeed implies $\eta \left ( \varphi,\psi \right )=(- 1)^{n_l(\varphi,\psi)-N/2}$ in (\ref{eq:overlap}). Let us compute $\langle \varphi \vert P_l \vert \chi \rangle$ where $\hat{P}_l$ is the permutation operator on the bond $l=(i,j)$. As shown on figure \ref{fig:boucles}, three cases can occur:
\begin{enumerate}
\item[{\it a.}]  The bond $l$ connects two sites lying at an odd distance on the same loop. The total number of loops of the overlap diagram remains unchanged and a sign change occurs: $\langle \varphi \vert P_l \vert \chi \rangle = - \langle \varphi \vert \chi \rangle $.
\item[{\it b.}]  The bond $l$ connects two distinct loops. Two loops are merged into a single one in the overlap diagram and no sign change occurs: $\langle \varphi \vert P_l \vert \chi \rangle = +2 \langle \varphi \vert \chi \rangle $.
\item[{\it c.}]  The bond $l$ connects two sites lying at an even distance on the same loop. One loop is created in the overlap diagram by slicing one loop into two and a no sign change occurs: $\langle \varphi \vert P_l \vert \chi \rangle = +1/2 \langle \varphi \vert \chi \rangle $.
\end{enumerate}

Obviously, the state $P_l \vert \chi \rangle$ violates the convention (\ref{eq:orientation}) since it is deduced by applying a 2-site permutation on a conventional state without being multiplied by the factor $\sigma(P_l)=-1$. The corresponding conventional state is thus  $\vert \psi \rangle = -\vert \chi \rangle$ and the three cases considered above become:
\begin{enumerate}
\item[{\it a.}]  $\langle \varphi \vert \psi \rangle = \langle \varphi \vert \chi \rangle $.
\item[{\it b.}]  $\langle \varphi \vert \psi \rangle = (-2) \langle \varphi \vert \chi \rangle$.
\item[{\it c.}]  $\langle \varphi \vert \psi \rangle = (-1/2) \langle \varphi \vert \chi \rangle$.
\end{enumerate}
The group structure of the permutations ensures that (i) the signs of all states (\ref{eq:orientation}) can be consistently generated from the reference state (\ref{eq:ref}) and (ii) a change $\Delta n_l$ in the number of loops comes with a sign change $(-1)^{\Delta n_l}$. This indeed implies $\eta \left ( \varphi,\psi \right )=(- 1)^{n_l(\varphi,\psi)-N/2}$.

\subsection{Operators and Generating function}
\label{subsec:generating} 

In this appendix, we introduce the generating function $\Z$. As explained in Section \ref{sec:expfuse}, the Hamiltonian and Overlap operators can be expanded in powers of $\alpha$ as linear superpositions
of dimer flipping processes~$\og$:
\begin{subequations}\label{eq:HOexp0}
\begin{align}
\Ham & = \sg \alpha^{2\ng} \hg \og, \label{eq:Hexp0} \\
\Ov & = \sg \alpha^{2\ng} \og. \label{eq:Oexp0}
\end{align}
\end{subequations}

Here $2\ng$ denotes the \textit{order} of the \textit{graph} $g$ representing the process. Note that $\og$ is a shortcut notation for an implicit summation over all the lattice of all plaquettes with the shape $g$. By definition $\hat{\omega}_0=\1$ and $h_0=0$.

The \textit{degree} $\dg$ of a graph $g$ is the number of its connected parts. By definition, if $\obare$ is a \textit{connected graph} then $\mathrm{dg}(\obare)=1$.
By convention, $\mathrm{dg}(\1)=0$ where $\1$ is the identity operator. For some examples, see Table \ref{tab:expansionorders}.

By introducing the $\Xg$ operators defined as,
\begin{equation}
\label{eq:Xg}
\Xg = \emuhg \og,
\end{equation}
such as $\muzero[{\dmu[\Xg]}]  = \hg \og$ and $\muzero[\Xg]  = \og$, one can express $\Ham$ and $\Ov$ as
\begin{subequations}\label{eq:HOexp2}
\begin{align}
\Ham & = \muzerobetaone[{\dmu[\Z]}], \label{eq:Hexp2} \\
\Ov  & = \muzerobetaone[\Z], \label{eq:Oexp2}
\end{align}
\end{subequations}
using the generating operator $\Z$ defined as,
\begin{equation}\label{eq:gen}
\Z (\alpha,\beta,\mu)= \sg \alpha^{2\ng} \beta^{\dg} \Xg.
\end{equation}

In turn, the effective Hamiltonian,
\begin{equation}\label{eq:Heff0}
\Heff = \Ov[-1/2] \, \Ham  \, \Ov[-1/2],
\end{equation}
is rewritten as,
\begin{equation}\label{eq:Heff1}
\Heff \eqmuzerobetaone \Z[-1/2] \left ( \dmu[\Z]  \right ) \Z[-1/2].
\end{equation}

\subsection{$\Heff$ as a function of $\logZ$}
\label{subsec:logZ}

The purpose of this Appendix is to explicit the transformation by which $\Heff$ given by \eqref{eq:Heff1} is rewritten
using only $\logZ$, see \eqref{eq:Heffcommutators}, as announced in \ref{sec:locality}

We use the general result,
\begin{equation}\label{eq:derive0}
\partial \left ( e^{F}  \right )= \int_{0}^{1} e^{(1-s)F} \left ( \partial F \right ) e^{s F}\,{\text{d}}s,
\end{equation}
for $ F = \logZ $ to express $\Heff$ as,

\begin{equation}\label{eq:Heff2}
\Heff \eqmuzerobetaone \int_{-\frac{1}{2}}^{\frac{1}{2}} \Z[-s] \left ( \dlogZ  \right ) \Z[s] \,{\text{d}}s.
\end{equation}

We now use the Hadamard formula
\begin{equation}
\label{eq:hadamard}
e^{A} B e^{-A} = e^{\halfcommute[A]} B \equiv B + \commute[A]{B} + \frac{1}{2!} \commute[A]{\commute[A]{B}} + \ldots
\end{equation}
with $A = s \logZ $ and $B = \dlogZ $ to rewrite \eqref{eq:Heff2} as,
\begin{equation}\label{eq:Heff3}
\Heff \eqmuzerobetaone \left ( \int_{-\frac{1}{2}}^{\frac{1}{2}} e^{-s \halfcommute[\logZ]} \,{\text{d}}s \right ) \dlogZ.
\end{equation}
Other equivalent forms are,
\begin{equation}\label{eq:Heff4}
\Heff \eqmuzerobetaone 2 \frac{\sinh \left ( \frac{1}{2} \halfcommute[\logZ] \right )}{\halfcommute[\logZ]} \dlogZ
\end{equation}
and
\begin{equation}\label{eq:Heff5}
\Heff \eqmuzerobetaone \sum_{p=0}^{\infty} \frac{1}{2^{2p}} \frac{\left ( \halfcommute[\logZ] \right )^{2p}}{\left ( 2 p + 1 \right )!} \dlogZ.
\end{equation}

\subsection{Cumulants}
\label{subsec:cumulants}

In this appendix we derive the expression of $\logZ$ entering in (\ref{eq:Heff3}), (\ref{eq:Heff4}) and (\ref{eq:Heff5}) as a $\beta$ expansion of non-commutative cumulants.

Let us introduce the collection of all $\alpha^{2\ng} \Xg$ of same degree $d$:
\begin{equation}\label{eq:Xd}
\cZ[d] = \sum_{\substack{g {\text{ such as}}\\  \dg=d}} \alpha^{2\ng} \Xg.
\end{equation}
$\cZ[1]$ involves \textit{all} simply connected graphs, $\cZ[2]$ all graphs of degree 2, etc and $\cZ[0]=\1$. Combining (\ref{eq:gen}) and (\ref{eq:Xd}) leads to
\begin{equation}
\label{eq:X}
\Z=\sum_{n=0}^\infty \beta^n \cZ[n],
\end{equation}
and
\begin{equation}\label{eq:logz0}
\logZ = -\sum_{n=1}^{\infty}\frac{1}{n} \left ( \halfacommute[-\frac{\Z-\1}{2}] \right )^n \1.
\end{equation}

Let us call $\Set[n]{m}$ the set of all sequences $s=\{s_i\}$ made of strictly positive integers $s_i$ such as $\text{Card}(s)=n$ and $\sum_{i \in s} s_i = m$. Notice that considering non-vanishing integer sequences, the minimal value of $\sum_{i \in s} s_i$ for $s \in \Set[n]{m}$ is $n$ and therefore $\Set[n]{m}$ is non-empty only for $m\geq n$. Moreover, $\Set[0]{m}$ is the empty set for all $m$. Using this notation,
\begin{equation}\label{eq:logz1}
\logZ = -\sum_{n=1}^{\infty} \sum_{m=n}^{\infty} \frac{(-1)^n}{n\,2^n} \beta^m \sum_{s \in \Set[n]{m}} \left ( \prod_{i=1}^{n} \halfacommute[{\cZ[s_i]}] \right ) \1,
\end{equation}
which, by inverting summations, can be written in turn as,
\begin{equation}\label{eq:logz2}
\logZ = \sum_{m=1}^\infty \beta^m \Zc[m],
\end{equation}
where the order-$m$ {\em non-commutative cumulant} $\Zc[m]$ is defined as,
\begin{equation}\label{eq:cn0}
\Zc[m]=-\sum_{n=1}^m \frac{(-1)^n}{n\,2^n}  \sum_{s \in \Set[n]{m}} \left ( \prod_{i=1}^{n} \halfacommute[{\cZ[s_i]}] \right ) \1,
\end{equation}
for $m>0$. By definition $\Zc[0]=\1$.

The explicit expressions of the first $\Zc[m]$ are:
\begin{align}\label{eq:cn_explicit0}
\Zc[1] &= \cZ[1] \\
\Zc[2] &= \cZ[2]-\frac{1}{4}\acommute[{\cZ[1]}]{\cZ[1]} \nonumber \\
\Zc[3] &= \cZ[3]-\frac{1}{2}\acommute[{\cZ[1]}]{\cZ[2]}+\frac{1}{12} \acommute[{\cZ[1]}]{\acommute[{\cZ[1]}]{\cZ[1]}} \nonumber \\
\Zc[4] &= \cZ[4]-\frac{1}{2}\acommute[{\cZ[1]}]{\cZ[3]}-\frac{1}{4}\acommute[{\cZ[2]}]{\cZ[2]} \nonumber \\
&+\frac{1}{6} \acommute[{\cZ[1]}]{\acommute[{\cZ[1]}]{\cZ[2]}} +\frac{1}{12} \acommute[{\cZ[2]}]{\acommute[{\cZ[1]}]{\cZ[1]}} \nonumber \\
&- \frac{1}{32} \acommute[{\cZ[1]}]{\acommute[{\cZ[1]}]{\acommute[{\cZ[1]}]{\cZ[1]}}} \nonumber
\end{align}

Interestingly, these expressions of $\Zc[m]$ in terms of $\cZ[n]$ can be inverted using the fact $\Z = \exp ( \ln (\Z))$. In the spirit of (\ref{eq:logz0}) and (\ref{eq:logz1}),
\begin{align}\label{eq:inv_logz0}
\Z &= \sum_{m=0}^{\infty}\frac{1}{m!} \left ( \halfacommute[\frac{\logZ}{2}] \right )^m \1 \\
  &= \sum_{m=0}^{\infty} \sum_{n=m}^{\infty} \frac{\beta^n}{m!\,2^m} \sum_{s \in \Set[m]{n}} \left ( \prod_{i=1}^{n} \halfacommute[{\cZ[s_i]}] \right ) \1 \\
  &= \sum_{n=0}^{\infty} \beta^n \sum_{m=0}^{n} \frac{1}{m!\,2^m} \sum_{s \in \Set[m]{n}} \left ( \prod_{i=1}^{n} \halfacommute[{\cZ[s_i]}] \right ) \1.
\end{align}

This last expression allows to identify,

\begin{equation}
\label{eq:cn_explicit}
\cZ[n] = \sum_{m=1}^{n} \frac{1}{m!\,2^m} \sum_{s \in \Set[m]{n}} \left ( \prod_{i=1}^{n} \halfacommute[{\Zc[s_i]}] \right ) \1.
\end{equation}

Here is the explicit form of the first terms,
\begin{align}\label{eq:cn_explicit1}
\cZ[1] &= \Zc[1] \\
\cZ[2] &= \Zc[2]+\frac{1}{4}\acommute[{\Zc[1]}]{\Zc[1]} \nonumber\\
\cZ[3] &= \Zc[3]+\frac{1}{2}\acommute[{\Zc[1]}]{\Zc[2]}+\frac{1}{24} \acommute[{\Zc[1]}]{\acommute[{\Zc[1]}]{\Zc[1]}} \nonumber\\
\cZ[4] &= \Zc[4]+\frac{1}{2}\acommute[{\Zc[1]}]{\Zc[3]}+\frac{1}{4}\acommute[{\Zc[2]}]{\Zc[2]}\nonumber \\
&+\frac{1}{12} \acommute[{\Zc[1]}]{\acommute[{\Zc[1]}]{\Zc[2]}} +\frac{1}{24} \acommute[{\Zc[2]}]{\acommute[{\Zc[1]}]{\Zc[1]}} \nonumber \\
&+ \frac{1}{192} \acommute[{\Zc[1]}]{\acommute[{\Zc[1]}]{\acommute[{\Zc[1]}]{\Zc[1]}}} \nonumber
\end{align}

\subsection{Diagrammatic notation}
\label{subsec:diagrammatic}

This Appendix is devoted to the introduction of a lattice-independent diagrammatic notation. Let us introduce the following diagrammatic notation :
\begin{equation}
\label{eq:diag_d}
\underbrace{\gd{NDisconnected}}_{n} = n! \, \cZ[n],
\end{equation}
and its cumulant counterpart,
\begin{equation}
\label{eq:diag_c}
\underbrace{\gdc{NDisconnected}}_{m} = m! \, \Zc[m],
\end{equation}

Defining, for any $s = \{ s_1,\ldots, s_m\} \in \Set[m]{n}$, the combinatorial factor
\begin{equation}
\label{eq:combinatorial}
\comb{s} = \frac{1}{m!} \frac{\left ( \sum_{p=1}^{m} s_p \right )!}{\prod_{p=1}^m s_p !},
\end{equation}
we rewrite equations (\ref{eq:cn0}) and (\ref{eq:cn_explicit}) as,
\begin{align}
\label{eq:zdiag1}
&\mean{\underbrace{\gdc{NDisconnected}}_{m}}=  \\
& \sum_{n=1}^{m} (-1)^{n-1} (n-1)! \sum_{s \in \Set[n]{m}} \comb{s} \mean{\underbrace{\gd{NDisconnected}}_{s_1} \times \ldots \times \underbrace{\gd{NDisconnected}}_{s_n}} \nonumber\\
&\mean{\underbrace{\gd{NDisconnected}}_{n}}=  \\
\label{eq:zdiag2}
& \sum_{m=1}^{n} \sum_{s \in \Set[m]{n}} \comb{s} \mean{\underbrace{\gdc{NDisconnected}}_{s_1} \times \ldots \times \underbrace{\gdc{NDisconnected}}_{s_m}} \nonumber,
\end{align}
where the brackets $\langle . \rangle$ notation stands for:
\begin{equation}
\label{eq:brackets}
\left \langle \prod_{i=1}^{n} \hat{o}_i \right \rangle = \frac{1}{2^n} \left ( \prod_{i=1}^{n} \left \{ \hat{o}_i, \right . \right ) \1.
\end{equation}

The diagrammatic counterparts of equations (\ref{eq:cn_explicit0}) and (\ref{eq:cn_explicit1}) are respectively
\begin{align}\label{eq:cn_explicit0_diag}
\mean{\gdc{1}} &=   0! \, \comb{1} \mean{\gd{1}} \\
\mean{\gdc{11}} &= 0! \, \comb{2} \mean{\gd{11}} \nonumber \\
&- 1! \, \comb{1,1}  \mean{\gd{1}\times\gd{1}} \nonumber \\
\mean{\gdc{111}} &= 0! \, \comb{3} \mean{\gd{111}} \nonumber \\
&- 1! \, \comb{1,2}  \mean{\gd{1}\times\gd{11}} \nonumber \\
&- 1! \, \comb{2,1}  \mean{\gd{11}\times\gd{1}} \nonumber \\
&+ 2! \, \comb{1,1,1}  \mean{\gd{1}\times\gd{1}\times\gd{1}}\nonumber \\
\mean{\gdc{1111}} &= 0! \, \comb{4} \mean{\gd{1111}} \nonumber \\
&- 1! \, \comb{1,3}  \mean{\gd{1}\times\gd{111}} \nonumber \\
&- 1! \, \comb{3,1}  \mean{\gd{111}\times\gd{1}} \nonumber \\
&- 1! \, \comb{2,2}  \mean{\gd{11}\times\gd{11}} \nonumber \\
&+ 2! \, \comb{1,1,2}  \mean{\gd{1}\times\gd{1}\times\gd{11}} \nonumber \\
&+ 2! \, \comb{1,2,1}  \mean{\gd{1}\times\gd{11}\times\gd{1}} \nonumber \\
&+ 2! \, \comb{2,1,1}  \mean{\gd{11}\times\gd{1}\times\gd{1}} \nonumber \\
&- 3! \, \comb{1,1,1,1}  \mean{\gd{1}\times\gd{1}\times\gd{1}\times\gd{1}}\nonumber
\end{align}
and
\begin{align}\label{eq:cn_explicit1_diag}
\mean{\gd{1}} &=  \comb{1} \mean{\gdc{1}} \\
\mean{\gd{11}} &= \comb{2} \mean{\gdc{11}} \nonumber \\
&+ \comb{1,1}  \mean{\gdc{1}\times\gdc{1}} \nonumber \\
\mean{\gd{111}} &= \comb{3} \mean{\gdc{111}} \nonumber \\
&+ \comb{1,2}  \mean{\gdc{1}\times\gdc{11}} \nonumber \\
&+ \comb{2,1}  \mean{\gdc{11}\times\gdc{1}} \nonumber \\
&+ \comb{1,1,1}  \mean{\gdc{1}\times\gdc{1}\times\gdc{1}}\nonumber \\
\mean{\gd{1111}} &= \comb{4} \mean{\gdc{1111}} \nonumber \\
&+ \comb{1,3}  \mean{\gdc{1}\times\gdc{111}} \nonumber \\
&+ \comb{3,1}  \mean{\gdc{111}\times\gdc{1}} \nonumber \\
&+ \comb{2,2}  \mean{\gdc{11}\times\gdc{11}} \nonumber \\
&+ \comb{1,1,2}  \mean{\gdc{1}\times\gdc{1}\times\gdc{11}} \nonumber \\
&+  \comb{1,2,1}  \mean{\gdc{1}\times\gdc{11}\times\gdc{1}} \nonumber \\
&+  \comb{2,1,1}  \mean{\gdc{11}\times\gdc{1}\times\gdc{1}} \nonumber \\
&+  \comb{1,1,1,1}  \mean{\gdc{1}\times\gdc{1}\times\gdc{1}\times\gdc{1}}\nonumber
\end{align}

\subsection{Fusion rules}
\label{subsec:fusion}

While the details of each fusion rule between diagrams (see equations \eqref{eq:fusexample} in Section \ref{sec:expfuse}) is lattice-specific, important and  general fusion properties can be described using the lattice-independent diagrammatic notation introduced in Appendix \ref{subsec:diagrammatic}. Indeed, as shown in Appendix \ref{sec:linkedcluster}, these lattice-independent fusion rules are sufficient conditions to demonstrate that any cumulant \eqref{eq:zdiag1} is connected, hence proving that $\Heff$ is local.

When computing products, diagrams can fuse together by producing {\em all possible} contractions of terms $\contraction[0.5ex]{}{\bgd{1}}{\ds}{\bgd{1}}\bgd{1}\ds\bgd{1}$ 
{\em between dinstinct blocks} $\gd{NDisconnected}\times\ldots\times\gd{NDisconnected}$. Fusions inside blocks $\contraction[0.5ex]{(\bgd{1}\ds}{\bgd{1}}{\bgd{Dot}}{\bgd{1}}(\bgd{1}\ds\bgd{1}\bgd{Dot}\bgd{1})$
are {\em not allowed}.

As as an example, let us apply these rules to the products appearing in $\mean{\gdc{11}}$ and $\mean{\gdc{111}}$. 

\begin{align}
\label{eq:contraction1}
\contraction{\mean{\gd{1}\times\gd{1}}=\mean{\gd{11}}+\langle}{\gd{1}}{\times}{\gd{1}}%
\mean{\gd{1}\times\gd{1}}&=\mean{\gd{11}}+\langle\gd{1}\times\gd{1}\rangle\\
&=\mean{\gd{11}}+\mean{\gd{2}}\nonumber
\end{align}

\begin{align}
\label{eq:contraction2}
&\mean{\gd{1}\times\gd{11}}=\mean{\gd{111}}\\
\contraction{+\langle(}{\bgd{1}}{)\times(}{\bgd{1}}
\contraction{+\langle(\bgd{1})\times(\bgd{1}\ds\bgd{1})\rangle+\langle(}{\bgd{1}}{)\times(\bgd{1}\ds}{\bgd{1}}
\contraction{+\langle(\bgd{1})\times(\bgd{1}\ds\bgd{1})\rangle+\langle(\bgd{1})\times(\bgd{1}\ds\bgd{1})\rangle+\langle(}{\bgd{1}}{)\times(}{\bgd{1}}
\contraction[2ex]{+\langle(\bgd{1})\times(\bgd{1}\ds\bgd{1})\rangle+\langle(\bgd{1})\times(\bgd{1}\ds\bgd{1})\rangle+\langle(}{\bgd{1}}{)\times(\bgd{1}\ds}{\bgd{1}}
&+\langle(\bgd{1})\times(\bgd{1}\ds\bgd{1})\rangle+\langle(\bgd{1})\times(\bgd{1}\ds\bgd{1})\rangle+\langle(\bgd{1})\times(\bgd{1}\ds\bgd{1})\rangle\nonumber\\
&=\mean{\gd{111}}+2\mean{(\bgd{2}\ds\bgd{1})}+\mean{\gd{3a}}\nonumber
\end{align}

\begin{align}
\label{eq:contraction3}
&\mean{\gd{11}\times\gd{1}}=\mean{\gd{111}}\\
\contraction{+\langle(}{\bgd{1}}{\ds\bgd{1})\times(}{\bgd{1}}
\contraction{+\langle(\bgd{1}\ds\bgd{1})\times(\bgd{1})\rangle+\langle(\bgd{1}\ds}{\bgd{1}}{)\times(}{\bgd{1}}
\contraction{+\langle(\bgd{1}\ds\bgd{1})\times(\bgd{1})\rangle+\langle(\bgd{1}\ds\bgd{1})\times(\bgd{1})\rangle+\langle(\bgd{1}\ds}{\bgd{1}}{)\times(}{\bgd{1}}
\contraction[2ex]{+\langle(\bgd{1}\ds\bgd{1})\times(\bgd{1})\rangle+\langle(\bgd{1}\ds\bgd{1})\times(\bgd{1})\rangle+\langle(}{\bgd{1}}{\ds\bgd{1})\times(}{\bgd{1}}
&+\langle(\bgd{1}\ds\bgd{1})\times(\bgd{1})\rangle+\langle(\bgd{1}\ds\bgd{1})\times(\bgd{1})\rangle+\langle(\bgd{1}\ds\bgd{1})\times(\bgd{1})\rangle\nonumber\\
&=\mean{\gd{111}}+2\mean{(\bgd{2}\ds\bgd{1})}+\mean{\gd{3a}}\nonumber
\end{align}

\begin{align}
\label{eq:contraction4}
&\mean{\gd{1}\times\gd{1}\times\gd{1}}=\mean{\gd{111}} \\
\contraction{+\langle(}{\bgd{1}}{)\times(}{\bgd{1}}
\contraction{+\langle(\bgd{1})\times(\bgd{1})\times(\bgd{1})\rangle+\langle(}{\bgd{1}}{)\times(\bgd{1})\times(}{\bgd{1}}
&+\langle(\bgd{1})\times(\bgd{1})\times(\bgd{1})\rangle+\langle(\bgd{1})\times(\bgd{1})\times(\bgd{1})\nonumber\\
\contraction{+\langle(\bgd{1})\times(}{\bgd{1}}{)\times(}{\bgd{1}}
\contraction{+\langle(\bgd{1})\times(\bgd{1})\times(\bgd{1})\rangle+\langle(}{\bgd{1}}{)\times(}{\bgd{1}}
\contraction{+\langle(\bgd{1})\times(\bgd{1})\times(\bgd{1})\rangle+\langle(\bgd{1})\times(}{\bgd{1}}{)\times(}{\bgd{1}}
&+\langle(\bgd{1})\times(\bgd{1})\times(\bgd{1})\rangle+\langle(\bgd{1})\times(\bgd{1})\times(\bgd{1})\nonumber\\
\contraction{+\langle(}{\bgd{1}}{)\times(}{\bgd{1}}
\contraction[2ex]{+\langle(}{\bgd{1}}{)\times(\bgd{1})\times(}{\bgd{1}}
\contraction[2ex]{+\langle(\bgd{1})\times(\bgd{1})\times(\bgd{1})\rangle+\langle(}{\bgd{1}}{)\times(\bgd{1})\times(}{\bgd{1}}
\contraction{+\langle(\bgd{1})\times(\bgd{1})\times(\bgd{1})\rangle+\langle(\bgd{1})\times(}{\bgd{1}}{)\times(}{\bgd{1}}
&+\langle(\bgd{1})\times(\bgd{1})\times(\bgd{1})\rangle+\langle(\bgd{1})\times(\bgd{1})\times(\bgd{1})\nonumber\\
\contraction{+\langle(}{\bgd{1}}{)\times(}{\bgd{1}}
\contraction{+\langle(}{\bgd{1}}{)\times(\bgd{1})\times(}{\bgd{1}}
\contraction[2ex]{+\langle(}{\bgd{1}}{)\times(\bgd{1})\times(}{\bgd{1}}
&+\langle(\bgd{1})\times(\bgd{1})\times(\bgd{1})\rangle\nonumber\\
&=\mean{\gd{111}}+3\mean{(\bgd{2}\ds\bgd{1})}+3\mean{\gd{3a}}+\mean{\gd{3b}}\nonumber
\end{align}

Combining (\ref{eq:contraction1}), (\ref{eq:contraction2}),(\ref{eq:contraction3}) and(\ref{eq:contraction4}) with 
(\ref{eq:cn_explicit0_diag}) leads to:
\begin{align}\label{eq:cn_explicit0_contracted}
\mean{\gdc{1}}   &= \mean{\gd{1}} \\
\mean{\gdc{11}}  &= -\mean{\gd{2}}\nonumber \\
\mean{\gdc{111}} &= 3\mean{\gd{3a}}+2\mean{\gd{3b}}\nonumber
\end{align}

\subsection{Linked cluster theorem}
\label{sec:linkedcluster}
In this section we demonstrate the results suggested in (\ref{eq:cn_explicit0_contracted}): {\em only connected diagrams contribute to $\mean{\gdc{NDisconnected}}$.}

The proof will be given in two steps. First, we show that $\mean{\gdc{NDisconnected}}$ can be reexpressed as (\ref{eq:unconstraint}) by relaxing the constraint on internal contractions. In the second part we use (\ref{eq:unconstraint}) to establish (\ref{eq:recursion4}) which shows that cumulants of any given order can be reexpressed with lower order cumulants hence providing a demonstration of the above mentioned result by direct induction.

\subsubsection{Relaxing internal contractions}
The demonstration uses the key fact that internal contractions
\begin{equation*}
\contraction[0.5ex]{(\bgd{1}\ds}{\bgd{1}}{\bgd{Dot}}{\bgd{1}}(\bgd{1}\ds\bgd{1}\bgd{Dot}\bgd{1})
\end{equation*}
are {\em not allowed} when evaluating $\mean{\gdc{NDisconnected}}$.

Let us relax this constraint and introduce $\mean{\gdc{NDisconnected}}^{*}$ whose definition
is the same as $\mean{\gdc{NDisconnected}}$ but evaluated with {\em at least one internal contraction}. Obviously, the sum 
\begin{equation}
\label{eq:Omega}
\hat{\Omega}_m=\mean{\underbrace{\gdc{NDisconnected}}_{m}}+\mean{\underbrace{\gdc{NDisconnected}}_{m}}^{*}
\end{equation}
corresponds to a fully unrestricted evaluation where both internal and external contractions are allowed. In that case each term of equation (\ref{eq:zdiag1}) produces the same set of terms by contraction since all partitions become equivalent when the distinction between internal and external contractions is suppressed. Thus, the corresponding weight can be obtained by the formal identification $\gd{NDisconnected} \leftrightarrow 1$ into equation (\ref{eq:zdiag1}) or, according to (\ref{eq:diag_d}), $\cZ[n] \leftrightarrow (1/n!)$ into equation (\ref{eq:X}). The later implies $\Z \leftrightarrow \exp(\beta)$ and $\logZ \leftrightarrow \beta$. This shows that $\hat{\Omega}_1 =\mean{\gd{1}}$ and $\hat{\Omega}_m = 0$ for $m>1$ which leads to,
\begin{equation}
\label{eq:unconstraint}
\mean{\underbrace{\gdc{NDisconnected}}_{m}}=
\begin{cases}
-\mean{\underbrace{\gdc{NDisconnected}}_{m}}^{*}& \text{if $m>1$}, \\
\mean{\gd{1}} & \text{for $m=1$}.
\end{cases}
\end{equation}

\subsubsection{Cumulant order reduction}

The proof of the result will be given by mathematical induction. We suppose the result to be true up to rank $m-1$.

For $m>1$,
\begin{multline}
\label{eq:recursion_0}
\mean{\underbrace{\gdc{NDisconnected}}_{m}}\\
\shoveleft{\quad=\sum_{n=1}^{m-1} (-1)^n (n-1)! \times} \\
\sum_{s \in \Set[n]{m}} \comb{s} \mean{\underbrace{\gd{NDisconnected}}_{s_1} \times \ldots \times \underbrace{\gd{NDisconnected}}_{s_n}}^{*}.
\end{multline}
Notice that the sum over $n$ has been truncated up to $m-1$ since the $n=m$ term contains only size-1 blocks and can obviously not produce any internal contraction. This equation can be made explicit as,
\begin{multline}
\label{eq:recursion1}
\mean{\underbrace{\gd{NDisconnected}}_{s_1} \times \ldots \times \underbrace{\gd{NDisconnected}}_{s_n}}^{*}\\
\shoveleft{\quad =-\mean{\underbrace{\gd{NDisconnected}}_{m}}}\\
\shoveleft{\quad  +\sum_{g_1=1}^{s_1} \ldots \sum_{g_n=1}^{s_n}  \sum_{\gamma_{1} \in \Set[g_1]{s_1}} \ldots \sum_{\gamma_{n} \in \Set[g_n]{s_n}}
\prod_{p=1}^{n} \frac{s_p !}{g_p ! \prod_{q=1}^{g_p} \gamma_{p}^{q} !} }  \\
\mean{
(\overset{\gamma_{1}^{1}}{\bgd{1f}}\ds\overset{\gamma_{1}^{2}}{\bgd{1f}}\bgd{Dot}\overset{\gamma_{1}^{g_1}}{\bgd{1f}})
\times \ldots \times (\overset{\gamma_{n}^{1}}{\bgd{1f}}\ds\overset{\gamma_{n}^{2}}{\bgd{1f}}\bgd{Dot}\overset{\gamma_{n}^{g_n}}{\bgd{1f}})}
\end{multline}
where $\overset{\gamma}{\bgd{1f}}$ denotes the result of all connected contractions between $\gamma$ operators $\bgd{1}$. Note that the presence of the first term on the r.h.s of equation (\ref{eq:recursion1}) takes into account that at least one internal contraction has to be performed.

For $\gamma_{p} \in \Set[g_p]{s_p}$ with $p=1,\ldots,n$ and $s \in \Set[n]{m}$ we have $\sum_{p=1}^{n} s_p = \sum_{p=1}^{n} \sum_{q=1}^{g_p} \gamma_{p}^{q}$ and thus,
\begin{equation}
\label{eq:combinatorialidentity}
\comb{s} \prod_{p=1}^{n} \frac{s_p !}{g_p ! \prod_{q=1}^{g_p} \gamma_{p}^{q} !} = \comb{\{g_1,\ldots,g_n\}} \comb{\gamma_1\cup\ldots\cup\gamma_n}.
\end{equation}

Using (\ref{eq:combinatorialidentity}) and combining (\ref{eq:recursion_0}) and (\ref{eq:recursion1}) leads to,
\begin{widetext}
\begin{align}
\label{eq:recursion2a}
&\mean{\underbrace{\gdc{NDisconnected}}_{m}} = \\
&\sum_{n=1}^{m-1} (-1)^n (n-1)! \sum_{s \in \Set[n]{m-1}} \sum_{g_1=1}^{s_1} \ldots \sum_{g_n=1}^{s_n}  \sum_{\gamma_{1} \in \Set[g_1]{s_1}} \ldots \sum_{\gamma_{n} \in \Set[g_n]{s_n}} 
\comb{\{g_1,\ldots,g_n\}} \comb{\gamma_1\cup\ldots\cup\gamma_n}  \mean{
(\overset{\gamma_{1}^{1}}{\bgd{1f}}\ds\overset{\gamma_{1}^{2}}{\bgd{1f}}\bgd{Dot}\overset{\gamma_{1}^{g_1}}{\bgd{1f}})
\times \ldots \times (\overset{\gamma_{n}^{1}}{\bgd{1f}}\ds\overset{\gamma_{n}^{2}}{\bgd{1f}}\bgd{Dot}\overset{\gamma_{n}^{g_n}}{\bgd{1f}})}, \nonumber
\end{align}
where summations over $s$ have been restricted to $s \in \Set[n]{m-1}$ to ensure that at least one internal contraction occurs. Next, a simple change of variables from $\gamma$ to $\mu$ and a sum inversion gives,
\begin{equation}
\label{eq:recursion2b}
\begin{split}
\mean{\underbrace{\gdc{NDisconnected}}_{m}} & = \sum_{n=1}^{m-1} (-1)^n (n-1)! \sum_{G=n}^{m-1}  \sum_{g \in \Set[n]{G}} \sum_{\mu \in \Set[G]{m}} \comb{g} \comb{\mu} \mean{
\underbrace{(\overset{\mu_1}{\bgd{1f}}\ds\overset{\mu_2}{\bgd{1f}}\bgd{Dot}\overset{\mu_{g_1}}{\bgd{1f}})}_{g_1}
\times \ldots \times \underbrace{(\bgd{1f}\ds\bgd{1f}\bgd{Dot}\overset{\mu_G}{\bgd{1f}})}_{g_n}}  \\
& = \sum_{G=1}^{m-1}\sum_{\mu \in \Set[G]{m}} \comb{\mu} \sum_{n=1}^{G} (-1)^n (n-1)!  \sum_{g \in \Set[n]{G}} \comb{g}  \mean{
\underbrace{(\overset{\mu_1}{\bgd{1f}}\ds\overset{\mu_2}{\bgd{1f}}\bgd{Dot}\overset{\mu_{g_1}}{\bgd{1f}})}_{g_1}
\times \ldots \times \underbrace{(\bgd{1f}\ds\bgd{1f}\bgd{Dot}\overset{\mu_G}{\bgd{1f}})}_{g_n}} \\
\end{split}
\end{equation}
\end{widetext}
This last expression involves explicitly the cumulants (\ref{eq:zdiag1}) and as a result, the order-$m$ cumulant can be expressed as a combination of lower order cumulants:
\begin{equation}
\label{eq:recursion3}
\mean{\underbrace{\gdc{NDisconnected}}_{m}} = -\sum_{G=1}^{m-1}\sum_{\mu \in \Set[G]{m}} \comb{\mu} \mean{
\underbrace{(\overset{\mu_1}{\bgd{1f}}\ds\overset{\mu_2}{\bgd{1f}}\bgd{Dot}\overset{\mu_{G}}{\bgd{1f}})_c}_{G}}
\end{equation}

The demonstration can be readily extended to the case where the operators in the l.h.s. of (\ref{eq:recursion3}) are arbitrary connected operators $\overset{\tau_p}{\bgd{1f}}$. With $\Gamma=\sum_{p=1}^{m} \tau_p$,
\begin{equation}
\label{eq:recursion4}
\mean{\underbrace{(\overset{\tau_1}{\bgd{1f}}\ds\overset{\tau_2}{\bgd{1f}}\bgd{Dot}\overset{\tau_{m}}{\bgd{1f}})_c}_{m}} = -\sum_{G=1}^{m-1}\sum_{\mu \in \Set[G]{\Gamma}} \comb{\mu} \mean{
\underbrace{(\overset{\mu_1}{\bgd{1f}}\ds\overset{\mu_2}{\bgd{1f}}\bgd{Dot}\overset{\mu_{G}}{\bgd{1f}})_c}_{G}},
\end{equation}
which concludes the recursion proof.

Using (\ref{eq:recursion3}), one easily recovers (\ref{eq:cn_explicit0_contracted}):
\begin{align}\label{eq:cn_explicit1_contracted}
\mean{\gdc{1}}   &= \mean{\gd{1}} \\
\mean{\gdc{11}}  &= -\comb{2} \mean{(\overset{2}{\bgd{1f}})_c} =  -\mean{\gd{2}} \\
\mean{\gdc{111}} &= -\comb{3} \mean{(\overset{3}{\bgd{1f}})_c} \\
								 &- \comb{1,2} \mean{(\overset{1}{\bgd{1f}}\ds\overset{2}{\bgd{1f}})_c} - \comb{2,1} \mean{(\overset{2}{\bgd{1f}}\ds\overset{1}{\bgd{1f}})_c} \nonumber\\
		             &= - \left ( \mean{\gd{3b}} + 3\mean{\gd{3a}} \right ) \nonumber \\
		             &+2\times\frac{3}{2} \left ( 2 \mean{\gd{3a}}+ \mean{\gd{3b}}\right ) \nonumber \\
		             &= 3\mean{\gd{3a}}+2\mean{\gd{3b}} \nonumber
\end{align}

\subsection{Fully connected diagrams}\label{sec:fullyconnected}

In practice the class of {\em fully connected diagrams},
\setlength{\abstractdiagwidth}{0.35mm}
\begin{equation*}
\bgd{NFullyConnected}
\end{equation*}
\setlength{\abstractdiagwidth}{0.23mm}
where each part $\bgd{1}$ is connected to the others plays a very important role when computing $\logZ$.  Indeed, they lead to the spatially most compact and prominent terms in $\Heff$, as explained in Section \ref{sec:locality}. The aim of this appendix is to compute the weight $w_m$ of this diagram in the order-$m$ cumulant:

\begin{equation}
\label{eq:maximal0}
\begin{split}
\mean{\underbrace{\gdc{NDisconnected}}_{m}} &= w_m \mean{{\overbrace{\left(\bgd{NFullyConnected}\right)}^{m}}_c}\\
&\quad + {\text{less connected terms}}
\end{split}
\end{equation}
Noticing that the way to connect two fully connected diagrams into a single fully connected one is unique, the recursion relation for $w_m$ is readily given by (\ref{eq:recursion3}),
\begin{equation}
\label{eq:maximal1}
w_m = -\sum_{p=1}^{m-1} w_p \sum_{\mu \in \Set[p]{m}} \comb{\mu}.
\end{equation}
Solving this recursive relation with $w_1=1$ leads to,
\begin{equation}
\label{eq:maximal2}
w_m=(-1)^{m-1} (m-1)!
\end{equation}
Finally, the contribution to $\logZ$, obtained by replacing (\ref{eq:maximal2}) into (\ref{eq:logz2}), is
\begin{equation}
\label{eq:maximal3}
\begin{split}
\logZ&=-\sum_{m=1}^{\infty} \frac{(-1)^m}{m!} \beta^m \mean{{\overbrace{\left(\bgd{NFullyConnected}\right)}^{m}}_c}\\
&\quad + {\text{less connected terms}}.
\end{split}
\end{equation}
}
\includepdf[pages={1-17,{}}]{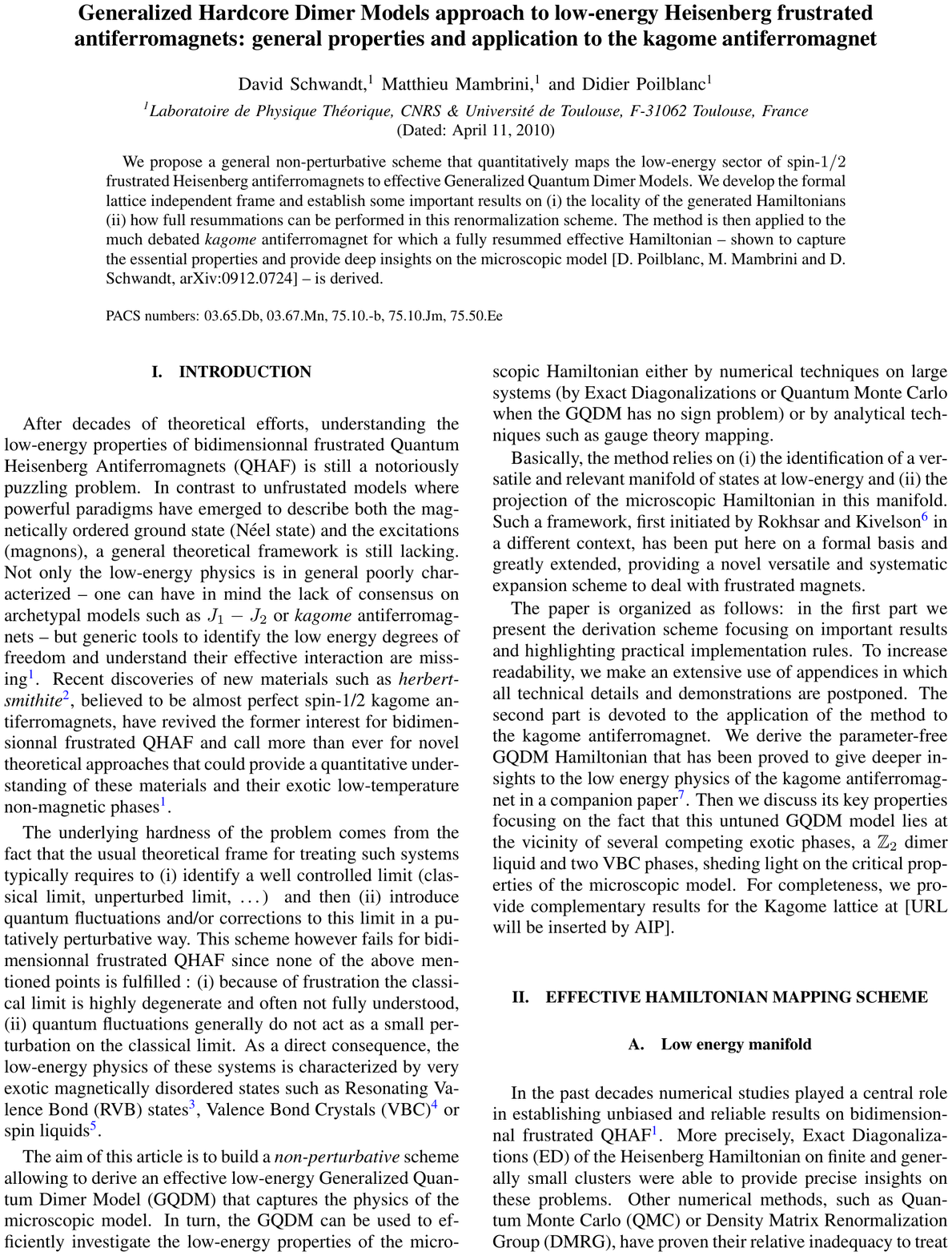}
\includepdf[pages={1-8}]{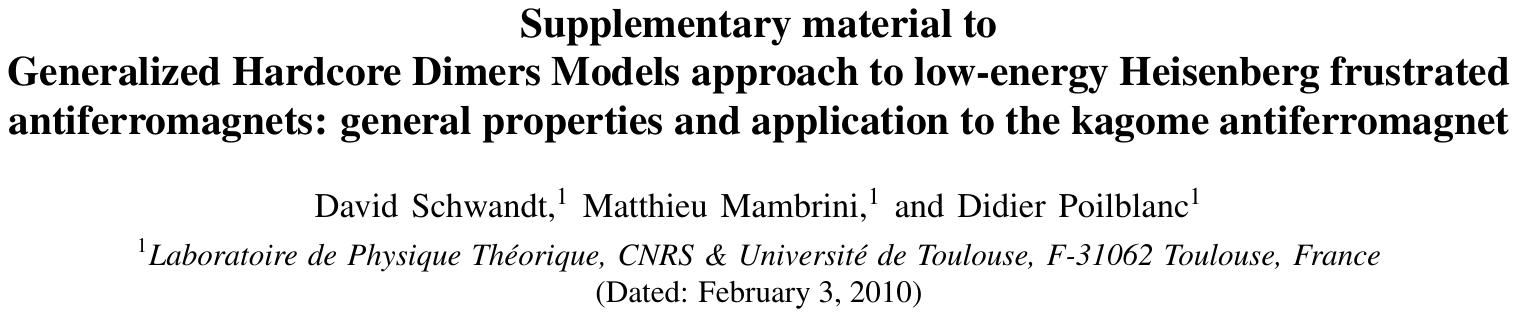}
%
%
\incl{%

\vfill
}

\newpage

\begin{thebibliography}{99}

\bibitem{j1j2} For reviews see e.g. {\it "Frustration in quantum antiferromagnets"}, H.J. Schulz, T.
Ziman and D. Poilblanc, in ``Magnetic systems with competing
interactions'', p120-160, Ed. H.T. Diep, World-Scientific,
Singapore (1994); {\it "Two-dimensional quantum antiferromagnets"}, G. Misguich \& C. Lhuillier, in "Frustrated spin systems'', Ed. H. T. Diep, World-Scientific (2005). 

\bibitem{herbertsmithite} M. Shores, E. Nytko, B. Bartlett, and D. Nocera,
J. Am. Chem. Soc. {\bf 127}, 13462 (2005); 
P. Mendels, F. Bert, M. A. de Vries, A. Olariu, A. Harrison, F.
Duc, J. C. Trombe, J. S. Lord, A. Amato and C. Baines, 
Phys. Rev. Lett. {\bf 98}, 077204 (2007); 
A. Olariu, P. Mendels, F. Bert, F. Duc, J. C. Trombe, M. A. de Vries, and A. Harrison,
Phys. Rev. Lett. {\bf 100}, 087202 (2008); 
A. Zorko, S. Nellutla, J. van Tol, L. C. Brunel, F. Bert, F. Duc, J.-C. Trombe, M. A. de Vries, A. Harrison and P. Mendels,
Phys. Rev. Lett. {\bf 101}, 026405 (2008).

\bibitem{anderson} P.W. Anderson, Science {\bf 235}, 1196 (1987).

\bibitem{vbc}
N. Read and S. Sachdev, Phys. Rev. B {\bf 42}, 4568 (1990);
N. Read and S. Sachdev, Phys. Rev. Lett. {\bf 66}, 1773 (1991);
S. Sachdev and N. Read, Int. J. of Modern Phys. B {\bf 5}, 219 (1991).

\bibitem{lee-et-al}  
M. Hermele, Y. Ran, P. A. Lee, and X.-G. Wen,
Phys. Rev. B 77, 224413 (2008).

\bibitem{rokhsar} D.S. Rokhsar and S.A. Kivelson, \prl {\bf 61}, 2376 (1988).

\bibitem{qdmkagome} D. Poilblanc, M. Mambrini and D. Schwandt, arXiv:0912.0724.

\bibitem{rumer} G.~Rumer, E.~Teller and H.~Weyl, Nachr. Ges. Wiss. Goettingen,
Math.-Phys. Kl. (No. 5), 499 (1932); see also H.N.~Temperley
and E.H.~Lieb, Proc. R. Soc. London, Ser. A {\bf 322}, 251 (1971);
R.~Saito, J. Phys. Soc. Jpn. {\bf 59}, 482 (1990).

\bibitem{hulthen} L.~Hulth\'en, Ark. Mat., Astron. Fys. {\bf 26}, 1 (1938); see also M.~Karbach, K.-H.~M\"utter, P.~Ueberholz, and H.~Kr\"oger, \prb {\bf 48}, 13666 (1993).

\bibitem{liang} S. Liang, B. Doucot and P. W. Anderson, \prl {\bf 61}, 365 (1988); see also A.~W.~Sandvik, \prl {\bf 95}, 207203 (2005).

\bibitem{J1J2J3NNVB} M.~Mambrini, A.~Lauechli, D.~Poilblanc \& F.~Mila \prb {\bf 74}, 144422 (2006).

\bibitem{kagomeNNVB}  M.~Mambrini and F.~Mila, Eur. Phys. J. B {\bf 17}, 651 (2000).

\bibitem{kagomeNNVBimpurity} S.~Dommange, M.~Mambrini, B.~Normand and F.~Mila, \prb {\bf 68}, 224416 (2003).

\bibitem{mambriniunp} M.~Mambrini (unpublished).

\bibitem{seidel} A.~Seidel, \prb {\bf 80}, 165131 (2009).

\bibitem{sutherland} B. Sutherland, Phys. Rev. B {\bf 37}, 3786 (1988).

\bibitem{zeng-elser} C.~Zeng and  Veit Elser, Phys.~Rev. B {\bf 51}, 8318 (1995). 

\bibitem{triangular-QDM} R. Moessner and S.~L.~Sondhi, Phys. Rev. Lett. {\bf 86}, 1881 (2001).

\bibitem{lecheminant} P.~Lecheminant, B.~Bernu, C.~Lhuillier, L.~Pierre and P.~Sindzingre, Phys.~Rev. B {\bf 56}, 
2521 (1997). 

\bibitem{sindzingre-99} P.~Sindzingre, G.~Misguich, C.~Lhuillier, B.~Bernu, L.~Pierre, Ch.~Waldtmann, and H.-U.~Everts,
Phys.~Rev.~Lett. {\bf 84}, 2953 (1999).

\bibitem{mila} F.~Mila, \prl {\bf 81}, 2356 (1998).

\bibitem{sindzingre-lhuillier}  P. Sindzingre and C. Lhuillier, Eur. Phys. Lett.  \textbf{88},  27009 (2009).

\bibitem{dmrg} H. C. Jiang, Z. Y. Weng, and D. N. Sheng, \prl {\bf 101} 117203 (2009).

\bibitem{leung} P.W.~Leung and Veit Elser, Phys.~Rev.~B {\bf 47},  5459 (1993). 

\bibitem{mg} C.K. Majumdar and D.K. Gosh, J. Math. Phys. {\bf 10}, 1388, (1969).

\bibitem{marston} J. B. Marston and C. Zeng, J.~Appl.~Phys. {\bf 69}, 5962 (1991).

\bibitem{maleyev} A. V. Syromyatnikov and S. V. Maleyev, Phys.~Rev.~B {\bf 66}, 132408~(2002).

\bibitem{auerbach} Ran Budnik and Assa Auerbach, Phys.~Rev.~Lett. {\bf 93}, 187205 (2004).

\bibitem{senthil} P. Nikolic and T. Senthil, Phys.~Rev. B {\bf 68}, 214415 ~(2003).

\bibitem{singh} R.R.P. Singh and D.A. Huse, Phys. Rev. B {\bf 76}, 180407 (2007).

\bibitem{misguich-sindzingre}
G. Misguich and P.~Sindzingre, J. Phys. Cond. Matt. {\bf 19}, 145202 (2007).

\bibitem{note-sign1}  The spectrum of the GQDM is invariant under a global change of sign of all kinetic processes around single hexagons.
This corresponds to a change between bosonic and fermionic conventions for the dimers as discussed, for the square lattice, 
in the second paper of Ref.~\protect\onlinecite{holons}.

\bibitem{note-sign2} Note that a bosonic convention has been used in Refs.~\protect\onlinecite{zeng-elser,qdmkagome} leading to an overall change of sign for all the
kinetic processes in equation (\protect\ref{eq:Hefffull}). 

\bibitem{note-zeng-elser} Ref.~\protect\onlinecite{zeng-elser} also investigates {\it numerically} all higher-order resonances on a small finite 36-site cluster (restricting to the relevant topological sector).

\bibitem{waldtmann} C.~Waldtmann, H.-U.~Everts, B.~Bernu, C.~Lhuillier, P.~Sindzingre, P.~Lecheminant, L.~Pierre, Eur. Phys. J. B {\bf 2}, 501 (1998).

\bibitem{j1j2j3dimer} 
For the same procedure on the frustrated quantum antiferromagnet on the square lattice
see A. Ralko, M. Mambrini and D. Poilblanc, Phys. Rev. B {\bf 80}, 184427 (2009).

\bibitem{misguich-RK}
G. Misguich, D. Serban, and V. Pasquier,
Phys. Rev. Lett. {\bf 89}, 137202 (2002); The GQDM defined here has no diagonal terms and equal amplitudes for all 32 kinetic processes
involving loops encircling a single hexagon. 

\bibitem{sachdev-QCP} C. Xu and S. Sachdev, Phys. Rev. B {\bf 79}, 064405 (2009).

\bibitem{triangular-condensation} A.~Ralko, M.~Ferrero, F.~Becca, D.~Ivanov, and F.~Mila, Phys. Rev. B 76, 140404 (2007). 

\bibitem{non-LGW} T. Senthil, Leon Balents, Subir Sachdev, Ashvin Vishwanath, and Matthew P. A. Fisher,
Phys. Rev. B {\bf 70}, 144407 (2004).

\bibitem{read} N.~Read and B.~Chakraborty, Phys. Rev. B {\bf 40}, 7133 (1989).

\bibitem{holons} For holon dynamics in a simple quantum dimer model on the square lattice see D. Poilblanc et al., \prb {\bf 74}, 014437 (2006); D. Poilblanc, \prl {\bf 100}, 157206 (2008).

\bibitem{spinons} For spinon dynamics in a simple quantum dimer model on the square lattice see A. Ralko, F. Becca and D.~Poilblanc, Phys.~Rev.~Lett. {\bf 101}, 117204 (2008).


\end{thebibliography}
\end{document}